\documentclass{amsart}
\usepackage{graphicx}
\usepackage{hyperref}
\usepackage{graphics} 
\usepackage{siunitx}
\usepackage{graphicx} 
\usepackage{hhline}
\usepackage{url,cite}
\usepackage{tabularx}
\usepackage{float}
\usepackage{caption}
\usepackage{lmodern}
\begin{document}
\title{Comparative Analysis of Finite Difference and Finite Element Method for Audio Waveform Simulation} 
\author{Juliette FLORIN}

\date{February 3\small{$^{\mathrm r \mathrm d}$}, 2025}
\thanks{Advisors: Massimiliano Todisco, Michele Panariello, and Nicholas Evans \\EURECOM \\ 
  Department of Digital Security \\ Campus SophiaTech \\ CS 50193 \\ 06904 Sophia Antipolis cedex 
  \\ FRANCE}

\begin{abstract}
In many industries, including aerospace and defense, waveform analysis is commonly conducted to compute the resonance of physical objects, with the Finite Element Method (FEM) being the standard approach. The Finite Difference Method (FDM) is seldom used, and this preference is often stated without formal justification in the literature. 
In this work, the accuracy, feasibility, and time of simulation of FEM and FDM are compared by simulating the vibration of a guitar string. Python simulations for both methods are implemented, and their results are compared against analytical solutions and experimental data. Additionally, FDM is applied to analyze the sound of a cycling bell to assess its reliability compared to a real cycling bell. Final results show that both FEM and FDM yield similar error margins and accurately predict the system's behavior. Moreover, the errors from FEM and FDM follow the same periodicity with a phase shift when varying the assumed analytical tension and without a phase shift when changing the time interval. However, FEM converges faster with increasing mesh complexity, whereas FDM demonstrates quicker computational performance and achieves stable solutions even with bigger time intervals. Despite this FDM is limited to simpler configurations and often demands extensive mathematical formulation, which can become cumbersome for intricate shapes. For example, modeling a hemispherical object using FDM results in significant simulation times and big calculations. In conclusion, while FDM may offer faster convergence and computation time in certain cases, FEM remains the preferred method in industrial contexts due to its flexibility, scalability, and ease of implementation for complex geometries. 
\end{abstract} 

\maketitle


\pagebreak
\section{Introduction} 
Accurate numerical modeling of sound-producing structures is fundamental in musical acoustics, audio engineering, virtual instrument synthesis, and many other domains. Among the available computational methods, the Finite Element Method (FEM) is commonly employed for simulating the vibrational behavior of complex structures, such as musical instruments \cite{kaselouris2022}. However, its effectiveness compared to alternative approaches, such as the Finite Difference Method (FDM), remains underexplored.

\subsection{Objective }
This study aims to compare the performance and accuracy of the FDM and the FEM in simulating audio waveforms for various physical phenomena. Specifically in the domain of predicting the resonant frequencies, mode shapes, and acoustic response of physical objects. The goal is to generate and analyze the sound waveforms produced by two chosen objects: a guitar string and a simple cycling bell. The project will investigate how the physical properties of these objects influence the behavior of both models and the resulting audio simulations. 

The results of this research will contribute to a deeper understanding of computational acoustics, which is used for theoretical and practical applications in musical instrument modeling, acoustic design, and digital sound synthesis. Moreover, it will assess the trade-off between FDM and FEM when choosing what method to use.

\subsection{Context and Motivation}
A wave is an oscillatory phenomenon (a quantity that varies over time) propagating through space. In acoustics, it refers to vibrations of matter, meaning a mechanical motion oscillating in a continuous medium, whether fluid or solid. Examples of waves include the vibration of a string, waves on the surface of water, sound propagating through air, etc. Each object, for instance, a hemispherical shell representing a cycling bell or a string of a guitar, when experiencing mechanical vibration, produces frequencies that correspond to specific resonances. 

Studying these vibrations requires solving partial differential equations that describe different vibration types, such as longitudinal, transverse, flexural, and torsional. For simpler systems like strings or beams, these equations include the second-order D'Alembert equation and the fourth-order Euler-Bernoulli beam equation. The vibrational behavior of hemispherical and string structures depends on natural frequencies, vibration modes, geometry, and material properties. Understanding these properties is crucial for engineering and scientific applications, with various methods available for analysis. 

In the industry today, FEM is mainly used  \cite{konate1986}. In many reports, it is also mentioned to be easier to compute and more efficient than the FDM \cite{kaselouris2022} \cite{raibaud2018}. The statements comparing the two are made without any references. In the context of acoustic waves and computing the waveform from a vibrating object, we would like to provide insight on both methods.

\subsection{Models and Theory}

At least two methods exist that can be used to model a physical phenomenon. Other methods are mainly derived cases from these general ones. On the one hand, there is the FEM used to model complex objects without little simplifications. On the other hand, the FDM has the same purpose but is also implemented when a solution to an equation is only reachable through approximations. Those two methods used in modeling physical phenomena should give the same answer as the theory. However, they are numerical methods; thus, the results will convey an error, no matter the model.

\subsection{Finite Element Method (FEM)}
The FEM is a numerical technique for solving differential equations, particularly useful for complex geometries and variable material properties \cite{manet2012}. The process involves transforming the strong form of the governing equations into a weak form, followed by discretization into finite elements. This results in the need to compute global system matrices, such as the mass and stiffness matrices.

\begin{figure}[H]
	\centering
	\includegraphics[width=0.75\linewidth]{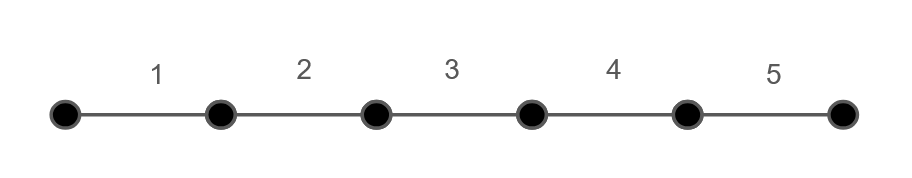}
	\caption{FEM representation example where the problem is split into multiple (numbered) elements}
	\label{fig:FEM}
\end{figure}

In FEM, division into small elements is done to approximate the solution using interpolation functions (shape functions) within each element. The unknowns, typically nodal displacements or field variables, are determined by assembling the contributions from individual elements into a global system of equations. We can see in the figure \ref{fig:FEM} an example of how to cut down a string into small elements. Each element has given length, weight, and properties, and must represent a small portion of the general object described.

\subsection{Finite Difference Method (FDM)}
Sometimes referred to as the Euler Method, the FDM is not widely adopted in the industry but remains highly effective when applied to well-understood problems. FDM is a numerical technique for solving differential equations by directly approximating derivatives at discrete points within the computational domain. Unlike the FEM, FDM does not require a weak formulation but instead discretizes the strong form of the governing equations.

\begin{figure}[H]
    \centering
    \includegraphics[width=0.75\linewidth]{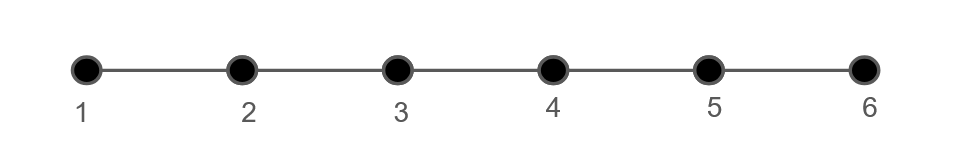}
    \caption{FDM representation example where the problem is split into multiple (numbered) nodes}
    \label{fig:FDM}
\end{figure}

This method approximates derivatives using finite difference approximations, including forward, backward, and central differences. These approximations transform the differential equations into algebraic equations, linking function values at discrete grid points. We can see in the figure \ref{fig:FDM} how small nodes can represent a string.

\section{Pressure to sound}
Once we have the amplitude in time of a given model, we can compute the waveform at a distance, given the pressure felt at that distance in time. In an acoustic simulation, the pressure generated by a source can be calculated using various physical parameters. If SI units are used for all quantities, the resulting acoustic pressure is expressed in Pascals (\si{\pascal}). However, to create a digital audio file, it is necessary to convert this acoustic pressure into amplitude and to quantize it and normalize it within the range \([-1, 1]\).

\subsection{Acoustic pressure model}

\begin{figure}[H]
    \centering
    \includegraphics[width=1\linewidth]{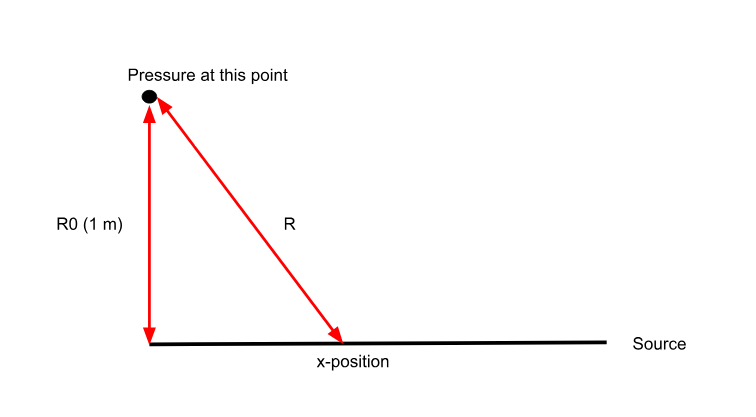}
    \caption{Representation of R during pressure calculation of a string}
    \label{fig:pressurer}
\end{figure}

The acoustic pressure \( p(t) \) generated at a distance R by a point source at position x is given in the equation \ref{eq:pressure_1}.

\begin{equation}
p(t) = \frac{\rho_0 c_0}{4 \pi} \frac{\partial u_R(x, t_r)}{\partial t}
\label{eq:pressure_1}
\end{equation}
where:

\begin{itemize}
    \item \( p(r, t) \) is the acoustic pressure in Pascals \si{\pascal}
    \item \( \rho_0 \) is the air density in \si{\kilogram\per\cubic\meter}
    \item \( c_0 \) is the speed of sound in air in \si{\meter\per\second}
    \item \( u_R(x, t_r) \) is the transverse displacement of the string in \si{\meter} with \(x\) being the position of the point source and \(t_r\) the time when the displacement happened
    \item \( \frac{\partial u_R}{\partial t_r} \) is the transverse velocity at the distance R from the point source in \si{\meter\per\second}
\end{itemize}
Let us note:
\begin{equation}
    [p(t)] = \si{\kilogram\per\cubic\meter} \cdot \si{\meter\per\second} \cdot \si{\meter\per\second} = \si{\pascal}
\end{equation}
The distance R, in \si{\meter}, appears when calculating \(\frac{\partial u_R(x, t_r)}{\partial t}\). In this project, the position of the calculated pressure is fixed at 1 \si{\meter} from one given point of the source. For each point \(x\) of the source, the corresponding \(R\) is calculated with the Pythagorean theorem. Figure \ref{fig:pressurer} illustrates R in the case of a string.
As the distance increases, the transverse velocity will decay by \(1/R\). Thus with \(\frac{\partial u_R(x, t_r)}{\partial t} = \frac{1}{R}\frac{\partial u(x, t-R/c_0)}{\partial t}\)  the calculated pressure is:

\begin{equation}
p(t)_{R} = \frac{\rho_0 c_0}{4 \pi R} \frac{\partial u(x, t- R/c_0)}{\partial t}
\end{equation}
Finally, to generate the pressure in time of a source, we add all the pressure generated at a given time by each point of a source. Each point of the source generates a given \(p(t)\) depending on its calculated \(R\) as mentioned in equation \ref{fig:pressurer}.
\subsection{Conversion of acoustic pressure to audio amplitude}

Several steps must be followed to convert the calculated acoustic pressure into an amplitude compatible with a digital audio recording. The first step is to normalize the acoustic pressure \( p(t) \) by a reference pressure \( p_{\text{max}} \), to avoid saturation in the audio file. The normalization formula is:

\begin{equation}
    A(t) = \frac{p(t)}{p_{\text{max}}}
\end{equation} 
where \( p_{max} \) is the maximum pressure in the simulation, and \( A(t) \) is the normalized amplitude within the range \([-1, 1]\).

Once the pressure is normalized, the amplitudes can be adapted to match digital audio formats. The goal is to map to the digital audio format. For example, for a 16-bit PCM audio file, the normalized amplitude \( A(t) \) can be converted to an amplitude within the range \([-32768, 32767]\) as follows:
\begin{equation}
    \text{Amplitude 16-bit} = \lfloor A(t) \cdot 32767 \rfloor
\end{equation}

Using Python, we can directly write the audio file in WAV format. Following these steps, the acoustic pressure calculated in a simulation can be converted into an appropriate audio amplitude, creating a digital audio file compatible with common standards (such as 16-bit PCM). This conversion ensures that the results from acoustic simulations can be easily analyzed or used in real-world audio applications.

\section{The guitar string without damping and without modulus of elasticity}
We are looking for the acoustic waves created when playing a guitar string without the flexion or the damping part of the equation \cite{alembert_eq}.  

\subsection{Model}

\begin{figure}[H]
    \centering
    \includegraphics[width=0.7\linewidth]{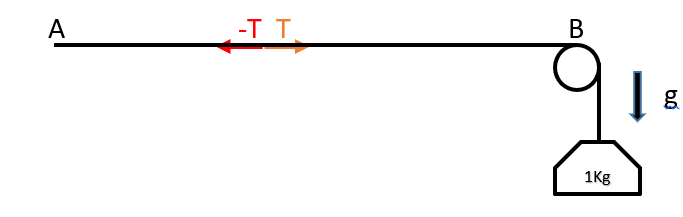}
    \caption{Example string for a mass of \( 1 \, \text{kg} \), the tension is \(T = 9.81 \si{\newton}\).}
    \label{fig:stringexample}
\end{figure}

Consider a string as seen in figure \ref{fig:stringexample} stretched along the horizontal axis between two fixed nodes, \( A \) and \( B \). To visualize the tension \( T \) of the string, imagine the string is attached at point \( A \) and pulled by a mass \( m \) via a pulley at point \( B \). The tension \( T \) is given by: \(T = m g\)  where \( g\) is 9.81 \si{\newton\kilogram}  is the acceleration due to gravity. When the string is at rest in equilibrium, the forces of tension in both directions cancel each other out (\( T \) and \( -T \)). The length of the string between its two fixed nodes is denoted as a product of A and B. The linear mass density string (mass per unit length) is represented by \( \mu \). 

Damping is not considered, as it does not affect the natural frequencies of the waveform. The modulus of elasticity is also neglected in the case of a nylon string, which means flexion is not included in the model. This simplification is valid for nylon due to its low stiffness, but not for metal strings, where the modulus of elasticity significantly contributes to flexural effects and must be accounted for \cite{alembert_eq} .

\subsection{Theory}

The wave equation for transverse vibrations of the string, derived by D'Alembert, is given by:
\begin{equation}
\frac{\partial^2 y(x, t)}{\partial t^2} = \frac{T}{\mu} \frac{\partial^2 y(x, t)}{\partial x^2}
\label{eq:transvibreation}
\end{equation}
where:

\begin{itemize}
    \item \( y(x, t) \) represents the transverse displacement of the string at position \( x \) and time \( t \)
    \item \( T \) is the tension of the string (in \si{\newton})
    \item \( \mu \) is the linear mass density of the string (in \si{\kilogram\per\meter})
    \item \( x \) is the spatial coordinate along the length of the string (in \si{\meter})
\end{itemize}
This equation \ref{eq:transvibreation} describes the string's motion under the influence of tension and its mass distribution.

The speed \( c \) (in \si{\meter\per\second}) of a transverse wave on the string is related to the tension \( T \) and the linear mass density \( \mu \) by the following equation:
\begin{equation}
     c = \sqrt{\frac{T}{\mu}}
     \label{eq:speed}
\end{equation}
This relationship shows that the wave speed increases with the tension and decreases with the mass per unit length of the string. The wave propagates faster for a string with higher tension, while a string with higher mass density will slow down the wave.

\subsection{Analytical solution}

The equation describing the transverse vibrations of a stretched string is the wave equation:
\begin{equation}
\frac{\partial^2 y(x,t)}{\partial t^2} = c^2 \frac{\partial^2 y(x,t)}{\partial x^2}
\end{equation}
where \( y(x,t) \) is the transverse displacement of the string at position \( x \) and time \( t \).

For a vibrating string fixed at both ends (i.e. \( y(0,t) = 0 \) and \( y(L,t) = 0 \)), we seek a solution in the form of standing waves. Thus, we assume a solution of the form:

\begin{equation}
y(x,t) = f(x) g(t)
\end{equation}
where \( f(x) \) is a spatial function that depends on the position \( x \) along the string, and \( g(t) \) is a temporal function that depends on time \( t \).
Substituting \( y(x,t) = f(x) g(t) \) into the wave equation, we get:

\begin{equation}
f(x) g''(t) = c^2 f''(x) g(t)
\end{equation}
Dividing both sides of this equation by \( f(x) g(t) \) gives:

\begin{equation}
\frac{g''(t)}{g(t)} = c^2 \frac{f''(x)}{f(x)}
\end{equation}
Since the left-hand side depends only on \( t \) and the right-hand side depends only on \( x \), both sides must be equal to a constant, which we will call \( -\omega^2 \), where \( \omega \) is the angular frequency. This gives us two separate equations:
\begin{align}
g''(t) &+ \omega^2 g(t) = 0
\label{eq:stringg}
\\
f''(x) &+ \frac{\omega^2}{c^2} f(x) = 0
\label{eq:stringf}
\end{align}
The general (real-valued) solution of equation \ref{eq:stringg} and \ref{eq:stringf} are respectively:
\begin{align}
    g(t) &= A \cos(\omega t) + B \sin(\omega t)
    \label{eq:g}
\\
    f(x) &= C \cos\left( \frac{\omega}{c} x \right) + D \sin\left( \frac{\omega}{c} x \right)
\end{align}
The boundary conditions are that the string is fixed at both ends:

\begin{equation}
    f(0) = 0 \quad \text{and} \quad f(L) = 0
\end{equation}
We apply these conditions to the solution for \( f(x) \).
This gives us that the quantized values for \( \omega \) are:
\begin{equation}
    \omega_n = \frac{n \pi c}{L} \quad \text{where} \quad n = 1, 2, 3, \dots
\end{equation}
We also have that at time 0, there is no speed:

\begin{equation}
 g'(0) = 0   
\end{equation}
With the boundary conditions, we get the result for \(g(t)\), according to the equation \ref{eq:g}:

\begin{equation}
   g(t) = A \cos(\omega t) 
\end{equation}
The general solution for the displacement of the string is thus an infinite sum of the vibrating modes:
\begin{equation}
    y(x,t) = \sum_{n=1}^{\infty} \left[ A_n \cos(\omega_n t) \right] \sin\left( \frac{n \pi}{L} x \right)
    \label{eq:generalsolstring}
\end{equation}
where the coefficients \( A_n \) are determined by the initial displacement.

\paragraph{Modes of vibration}
The angular frequency is related to the frequency and gives us all the superposing modes in the final waveform. This tells us what the fundamental frequency and the harmonics will be\cite{joussen1990} \cite{martin1989}.  
\begin{equation}
    \omega_n = 2\pi f_{n} = \frac{n \pi c}{L} \quad 
\end{equation}
The fundamental wavelength of vibration, the basic mode, corresponds to the string having two vibration nodes located at the fixed point. The length of the string corresponds to half the wavelength. Hence, the fundamental frequency of vibration is given by:
\begin{equation}
    f_1 = \frac{1}{2L} \sqrt{\frac{T}{\mu}}
    \label{eq:analyticalfunda}
\end{equation}
where:
\begin{itemize}
    \item \( L \) is the length of the string (in \si{\meter})
\end{itemize}
The harmonics of this fundamental will be \(nf_1\) with \(n\) an integer.
The harmonics depend on where the string is plugged. D'Alembert's wave equation for transverse vibrations of a plucked string describes the string's motion under tension. The fundamental frequency of vibration is determined by the tension of the string, the linear mass density, and the length of the string. This theory is widely used to understand the behavior of musical strings.

\subsection{Parameters for a nylon string B3 (247 Hz) }
The parameters for the nylon string chosen are\cite{polisano2016} \cite{td_ondes_mecaniques2017} \cite{derveaux2002}: 
\begin{itemize}
    \item \( L = 0.65 \ \)\si{\meter}
    \item\( \mu = 0.000582\) \si{\kilogram\per\meter}
    \item\( T = 60 \, \text{N} \)
\end{itemize}
Using the wave speed formula (equation \ref{eq:speed}) for a string and the fundamental frequency calculation (equation \ref{eq:analyticalfunda}), we calculate the wave speed to \(321\) \si{\meter\per\second}and the fundamental frequency to 247 \si{\hertz}

\subsection{Finite Difference Method}
We start by using FDM.
\subsubsection{Theory}
In this method, derivatives and second derivatives are replaced by their estimates:
\begin{align}
\frac{\partial u}{\partial x} &\approx \frac{u(x + \delta_x, t) - u(x, t)}{\delta_x}, \\
\frac{\partial u}{\partial t} &\approx \frac{u(x, t + \delta_t) - u(x, t)}{\delta_t}, \quad \\
\frac{\partial^2 u}{\partial x^2} &\approx \frac{u(x + \delta_x, t) - 2u(x, t) + u(x - \delta_x, t)}{\delta_x^2}, \\
\frac{\partial^2 u}{\partial t^2} &\approx \frac{u(x, t + \delta_t) - 2u(x, t) + u(x, t - \delta_t)}{\delta_t^2}. \quad 
\end{align}
Here, $\delta_x$ and $\delta_t$ are tiny intervals in space and time, respectively.
We discretize the space in length $x$ and time $t$, and rewrite D'Alembert's equation:
\begin{align}
&\frac{u(t + \delta_t, x) - 2u(t, x) + u(t - \delta_t, x)}{\delta_t^2}
\notag  \\
&- c^2 \frac{u(t, x + \delta_x) - 2u(t, x) + u(t, x - \delta_x)}{\delta_x^2} 
= 0
\label{eq:FDMrewrite}
\end{align}
By setting \(\gamma = \frac{c^2 \delta_t^2}{\delta_x^2}\) in equation \ref{eq:FDMrewrite},
we get:
\begin{equation}
    u(t + \delta_t, x) = \gamma \big(u(t, x - \delta_x) + u(t, x + \delta_x)\big) + 2(1-\gamma) u(t, x) - u(t - \delta_t, x)
\end{equation}
We divide the string of length $L$ into $N-1$ segments of length \(\delta_x = \frac{L}{N-1}\), 
and the observation time $A$ into $K$ intervals of duration \(\delta_t = \frac{A}{K-1}\). 
We study vibrations at the corresponding node and instants:
\begin{equation}
    t_k = (k-1) \delta_t, \quad x_n = (n-1) \delta_x, \quad u_{n,k} = u(x_n, t_k). \nonumber
\end{equation}

For a guitar string, the initial deformation is due to the plucking of the string at a position $x_p$. This can be expressed as:
\begin{equation}
    u(x, 0) =
\begin{cases} 
h \cdot \frac{x}{x_p}, & \text{if } 0 \leq x \leq x_p, \\
h \cdot \frac{L-x}{L-x_p}, & \text{if } x_p \leq x \leq L,
\end{cases}
\end{equation}
where $h$ is the amplitude of the pluck, $L$ is the length of the string, and $x_p$ is the position of the pluck. The initial velocity is zero everywhere:
\begin{equation}
    \frac{\partial u}{\partial t}(x, 0) = 0, \quad \forall x \in [0, L]
\end{equation}
In discrete notation, the initial conditions become:
\begin{align}
    u_{i,0} &=
\begin{cases} 
h \cdot \frac{x_i}{x_p}, & \text{if } 0 \leq x_i \leq x_p \\
h \cdot \frac{L-x_i}{L-x_p}, & \text{if } x_p \leq x_i \leq L
\end{cases}
\label{eq:inistring1}
\\ u_{i,1} &= u_{i,0}, \quad \forall i \in [0, N]
\label{eq:inistring2}
\end{align}
For a guitar string fixed at both ends, the boundary conditions are for all t:
\begin{equation}
    u(0, t) = u(L, t) = 0
\end{equation}
In discrete notation, this becomes:
\begin{equation}
    u_{0,j} = u_{N,j} = 0, \quad \forall j \in [0, K]
\label{eq:bondstring}
\end{equation}
\subsubsection{Computation and result}

\begin{figure}[H]
    \centering
    \includegraphics[width=1\linewidth]{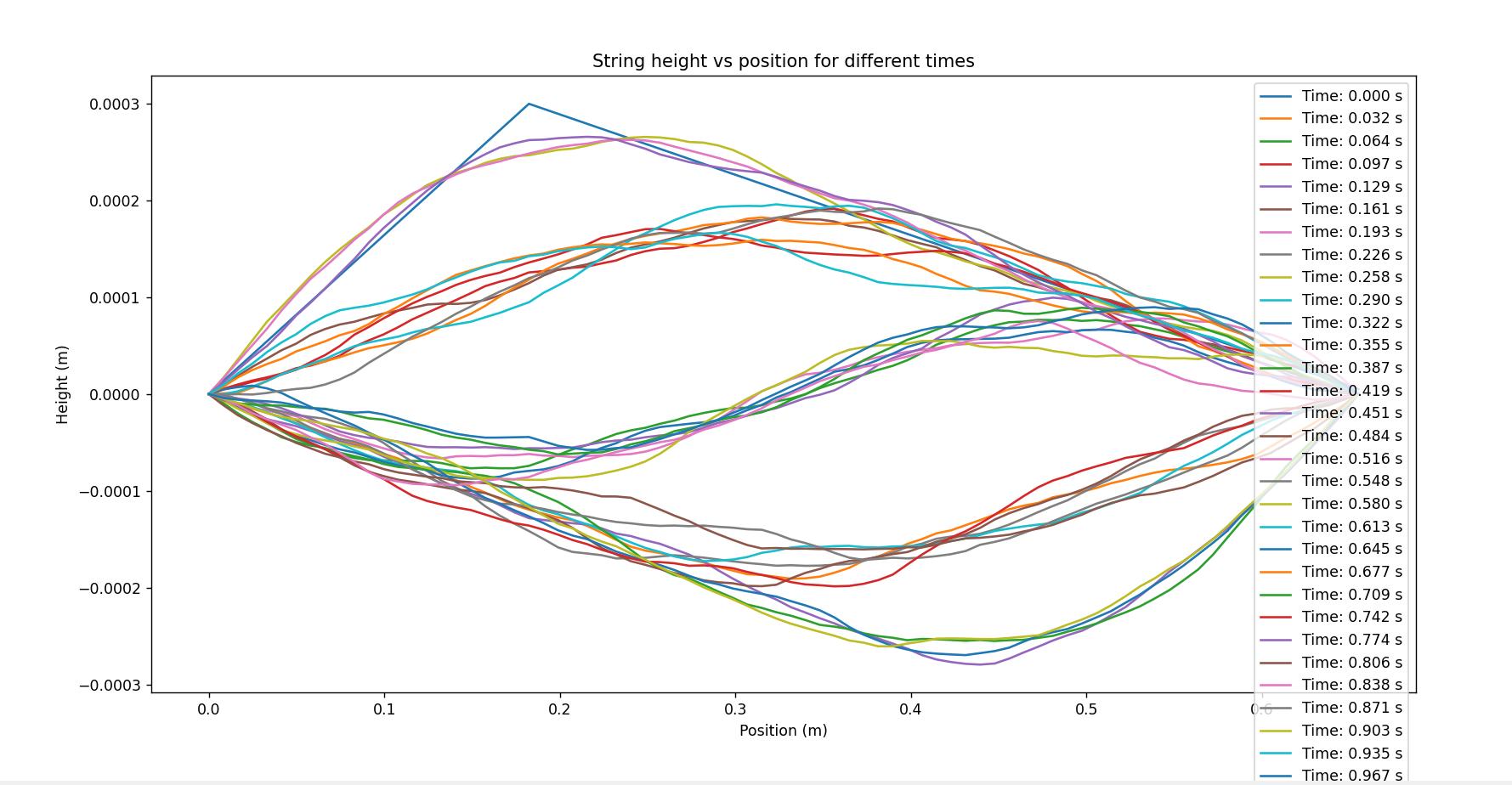}
    \caption{Shape in time of a nylon string B3 (247 Hz) plucked at 0.18 m for one second with 81 nodes (\(N = 80\))}
    \label{fig:Stringmage}
\end{figure}

\begin{figure}[H]
    \centering
    \includegraphics[width=1\linewidth]{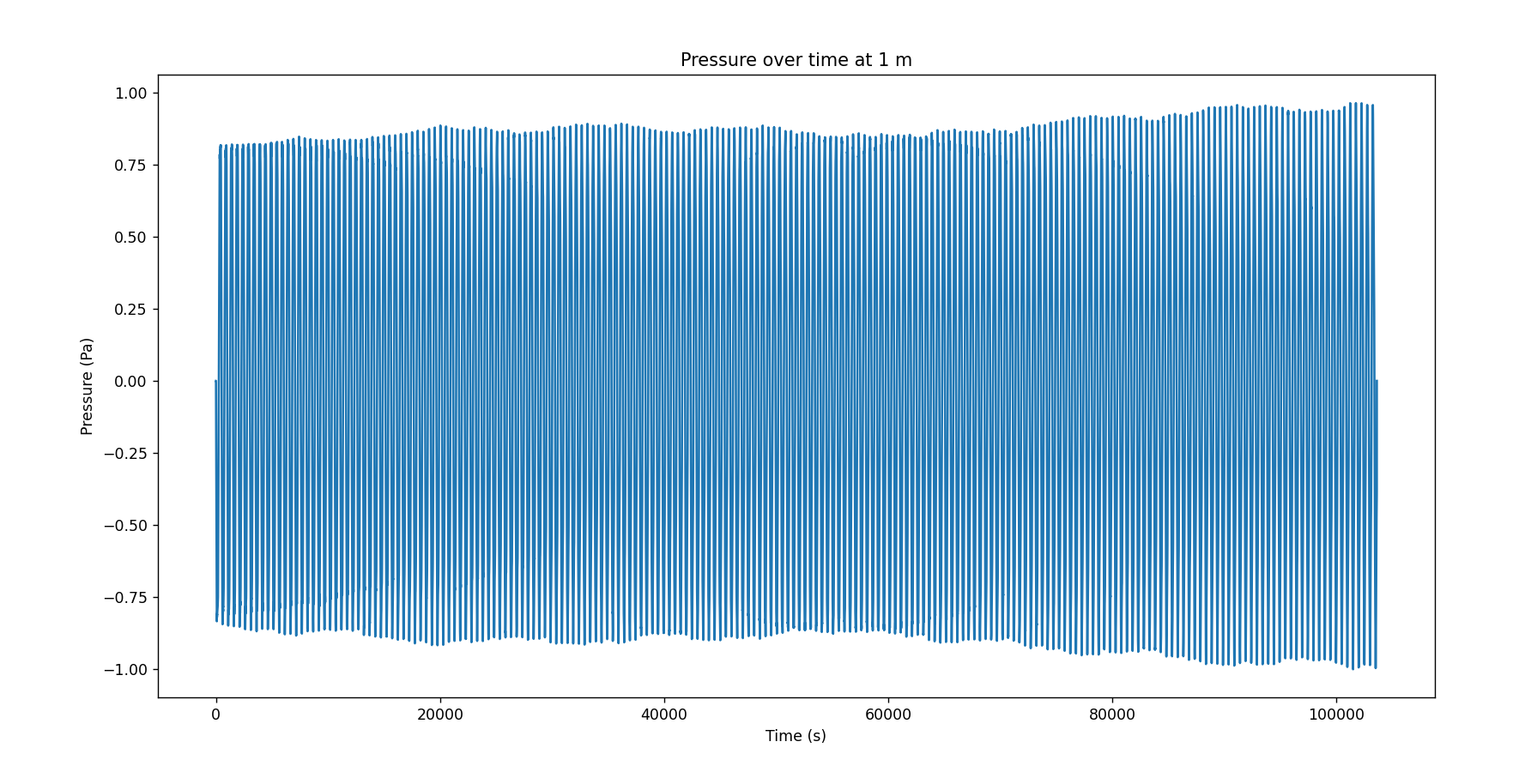}
    \caption{Pressure in time for a nylon string B3 (247 Hz) for one second with 81 nodes}
    \label{fig:pressureresult}
\end{figure}

\begin{figure}[H]
    \centering
    \includegraphics[width=1\linewidth]{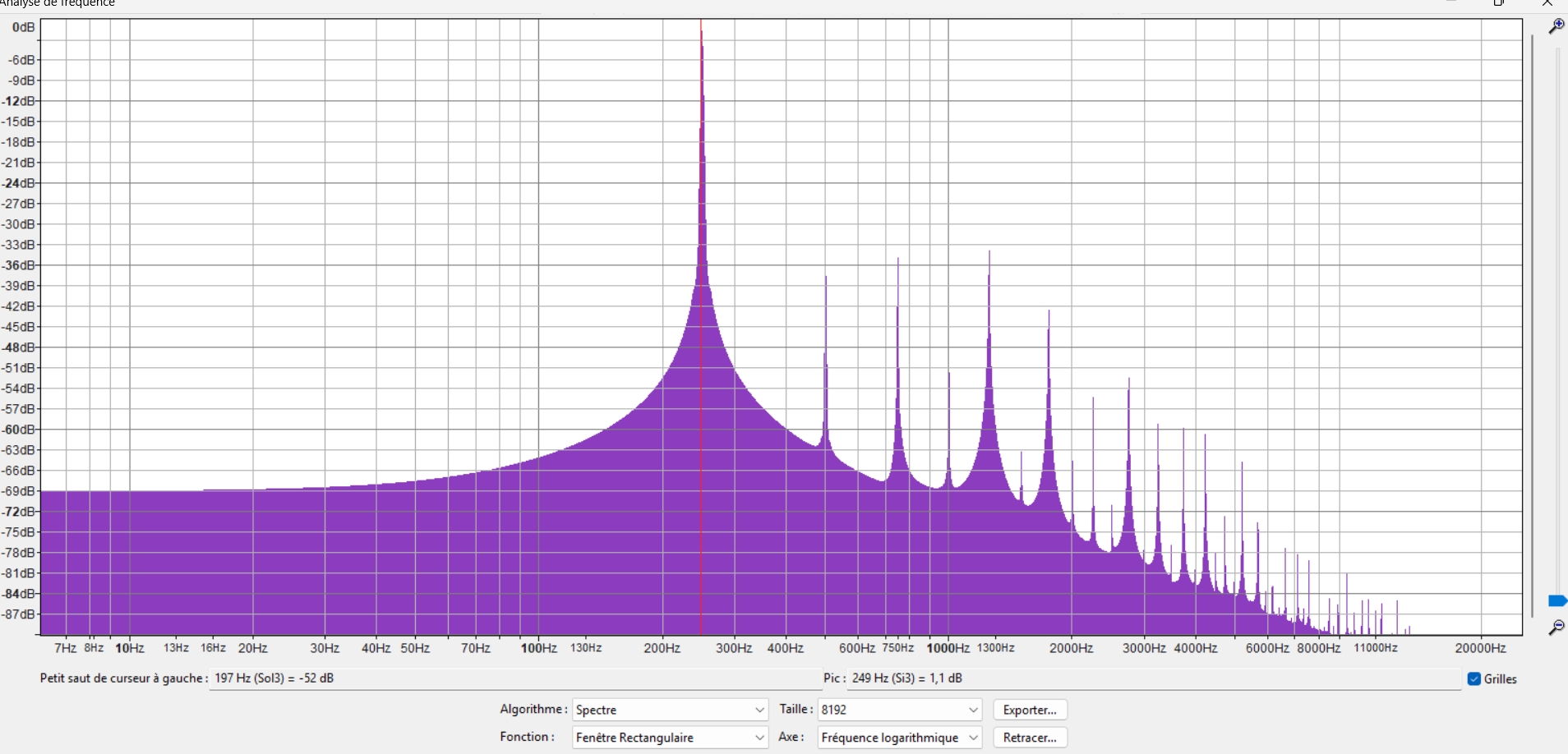}
    \caption{FFT of the signal of the guitar}
    \label{fig:fftresult}
\end{figure}

Figure \ref{fig:Stringmage} represents the string changing over time after being plucked. Figure \ref{fig:pressureresult} shows the pressure felt at 1\si{\meter} of the string over time. This will be used to create the waveform. Figure \ref{fig:fftresult} represents the Fast Fourier Transform (FFT) of the waveform of the sound of a plucked guitar string. The spectrum observed contains a large quantity of harmonics. Noticeably, the relative amplitudes of these harmonics depend on where the string is plucked. This phenomenon can be explained by the physics of standing waves and the resulting interaction between the fundamental frequency and its harmonics. For example, if you pluck the string at half its length (i.e. at its midpoint), the even harmonics are absent. This relationship can be understood by considering the Fourier series representation of the wave in equation \ref{eq:generalsolstring}. The position where the string is plucked determines which harmonics are most effectively excited, as different positions correspond to different initial conditions for the sinusoidal components of the wave. In half the distance, the sinusoid is null for the even harmonics.


\subsection{Finite Element Method}

The weak form of the wave equation is given as \cite{manet2012} \cite{deghboudj2018} \cite{deghboudj2023}:
\begin{equation}
    M \ddot{U}(t) + K U(t) = 0
\end{equation}
where:
\begin{itemize}
    \item \(M\): Global mass matrix (from inertia term),
    \item \(K\): Global stiffness matrix
    \item \(U(t)\): Vector of nodal displacements,
    \item \(\ddot{U}(t)\): Vector of nodal accelerations.
\end{itemize}
The goal is to compute \(U^{n+1}\), the displacement at the next time step, given the current displacements (\(U^n\), ) and displacements at the previous time step (\(U^{n-1}\)).

To discretize the wave equation over time, we approximate the second derivative \(\ddot{U}(t)\) using the central difference scheme:
\begin{equation}
    \ddot{U} \approx \frac{U^{n+1} - 2U^n + U^{n-1}}{\Delta t^2}
\end{equation}
Substitute this into the weak form:
\begin{equation}
    \frac{M}{\Delta t^2}(U^{n+1} - 2U^n + U^{n-1}) + K U^n = 0
\end{equation}
Simplify for \(U^{n+1}\):
\begin{equation}
    U^{n+1} = 2U^n - U^{n-1} - \Delta t^2 M^{-1} K U^n
\end{equation}
with:
\begin{itemize}
    \item \(2U^n\): The current displacement contributes the most to the next time step.
    \item \(-U^{n-1}\): The inertia (past displacement) opposes changes.
    \item \(-\Delta t^2 M^{-1} K U^n\): The stiffness and mass interaction modifies the displacement over time.
\end{itemize}
Now that we have \(U^{n+1}\) depending on previous \(U\) we can compute M and K. To do that we fist need to compute their local matrix. To compute the local mass matrix (\(M_{local}\)) and local stiffness matrix (\(K_{\text{local}}\)) for a single finite element in a 1D string using the FEM, follow these steps:

For a 1D element with linear shape functions, we define:
\begin{equation}
    N_1(x) = \frac{x_{e+1} - x}{\Delta x}, \quad N_2(x) = \frac{x - x_e}{\Delta x}
\end{equation}
where:
\begin{itemize}
    \item \(x_e\) and \(x_{e+1}\) are the positions of the two nodes of the element,
    \item \(\Delta x = x_{e+1} - x_e\) is the length of the element.
\end{itemize}

The mass matrix \(M_{\text{local}}\) is derived from the term:
\begin{equation}
    \int_{x_e}^{x_{e+1}} \rho N_i(x) N_j(x) \, dx
\end{equation}
For a two-node linear element (\(i, j = 1, 2\)), the result is:
\begin{equation}
    M_{\text{local}} = \frac{\rho \Delta x}{6}
    \begin{bmatrix}
    2 & 1 \\
    1 & 2
    \end{bmatrix}
\end{equation}
\begin{itemize}
    \item The diagonal terms (\(2\)) represent the inertia contribution of each node.
    \item The off-diagonal terms (\(1\)) represent the coupling between the two nodes.
\end{itemize}

The stiffness matrix \(K_{\text{local}}\) is derived from the terms in equation \ref{eq:stiffness} \cite{manet2012} \cite{deghboudj2018} \cite{deghboudj2023}.
\begin{equation}
    \int_{x_e}^{x_{e+1}} T \frac{\partial N_i}{\partial x} \frac{\partial N_j}{\partial x} \, dx
    \label{eq:stiffness}
\end{equation}
For linear elements, the derivatives of the shape functions are constant:
\begin{equation}
    \frac{\partial N_1}{\partial x} = -\frac{1}{\Delta x}, \quad \frac{\partial N_2}{\partial x} = \frac{1}{\Delta x}
\end{equation}
Substitute these into the integral:
\begin{equation}
    K_{\text{local}} = \frac{T}{\Delta x}
\begin{bmatrix}
1 & -1 \\
-1 & 1
\end{bmatrix}
\end{equation}
\begin{itemize}
    \item The diagonals with the value 1 represent the stiffness contribution at each node.
    \item The off-diagonal terms with the value \(-1\) represent the coupling between the two nodes.
\end{itemize}

The global mass matrix (\(M\)) and global stiffness matrix (\(K\)) are assembled by summing the contributions from all elements. Each local matrix is added to the global matrix by mapping its nodes (degrees of freedom) to the global indices. For a string that is fixed at both ends (\(u(0, t) = u(L, t) = 0\)), the boundary conditions are applied as follows:
\begin{itemize}
    \item Modify the global mass matrix (\(M\)) and stiffness matrix (\(K\)) such that the rows and columns corresponding to the boundary nodes are set to zero.
    \item Set the diagonal entries for the boundary nodes to 1, ensuring their displacements remain zero.
\end{itemize}
When implementing the global matrices in Python we can write it:
\[
M[0, :] = 0, \quad M[-1, :] = 0, \quad M[0, 0] = 1, \quad M[-1, -1] = 1
\]
\[
K[0, :] = 0, \quad K[-1, :] = 0, \quad K[0, 0] = 1, \quad K[-1, -1] = 1
\]
This enforces the fixed displacement at the boundaries while keeping the system consistent. We now put this in a Python code and simulate it.

Now that we have the recursion terms, \(M\) and \(K\), the algorithm used in the code will be: 
\begin{enumerate}
    \item Assemble the global mass and stiffness matrices:
    \begin{itemize}
        \item \(M\): coming from the weak form of the inertia term.
        \item \(K\): coming from the weak form of the elastic term.
    \end{itemize}
    These are computed for all elements and assembled into the global system.

    \item Boundary conditions: Apply boundary conditions to \(M\) and \(K\), such as fixing the ends of the string (displacement = 0).

    \item Iterative update: Use the update formula:
    \begin{equation}
        U^{n+1} = 2U^n - U^{n-1} - \Delta t^2 M^{-1} K U^n
    \end{equation}
    Compute \(U^{n+1}\) at every time step, starting with initial conditions:
    \begin{itemize}
        \item \(U^0\): Initial displacement (e.g., triangular shape from the pluck),
        \item \(U^1\): First step, often approximated as \(U^1 = U^0\).
    \end{itemize}

\end{enumerate}

\section{Comparing FEM and FDM guitar string}
With the implementation of both the FEM and the FDM in Python, the objective is to conduct a comparative analysis of the two approaches using various performance metrics. By systematically varying different parameters for each model, as detailed in the following sections, the waveform was computed for both FEM and FDM. Additionally, the expected theoretical frequencies (fundamental and harmonics) were calculated to be used as a reference. Using \textit{Audacity}, each fundamental frequency and its harmonics were recovered. It was achieved using a rectangular FFT of 4096 in size. \label{info:windowchoice} For all the FFT in this paper a rectangular window was chosen with a size of 4096. The different types of windows like the Bartlett window, Hamming window, rectangular window, Hann window, Blackman window, Blackman-Harris window, Welch window, and different Gaussian windows, were tried, and the rectangular window gave the best resolution of amplitude and frequency. The size of the window was then chosen according to the sampling rate. The goal was to keep the same window size along all the paper; thus, the window size was big enough to minimize the FFT error but small enough to show the frequencies when testing the impact of time interval. The relative error of the frequencies for each method was then computed and plotted across different scenarios, allowing for an evaluation of the accuracy and reliability of FEM and FDM in approximating the expected results.

\subsection{Variation of applied tension}
The tension is directly related to the frequency and the speed through an analytic formula as seen in equation \ref{eq:analyticalfunda}. This is true for the fundamental frequency but also for the harmonics as they are related to the fundamental by \(nf_1\) (named analytic frequencies in the following). To compare both methods, we are setting all the parameters and looking at the effect of a changing tension on the error of the measured frequencies compared to the analytical frequencies.
\begin{itemize}
    \item Length of the string: \( L = 0.655 \) \si{\meter}
    \item Position of plucking: half its length, \(0.3275 \si{\meter}\)
    \item Number of nodes the string is split into: \( N = 80 \)
    \item Time interval: \( \Delta t = 1 \times 10^{-5} \) \si{\second}
    \item Number of time points: \( N_t = 1 \times 10^{5} \)
    \item Total simulation time: \( T = 1 \) \si{\second}
    \item Amplitude of plucking: \( A = 3 \times 10^{-4} \) \si{\meter}
    \item Linear mass density: \( \mu = 4.30 \times 10^{-4} \)\si{\kilogram\per\meter}
\end{itemize}
40 waveforms were computed for 40 different tension values between 42 $N$ and 61.5 $N$ every 0.5 step. The values were taken to ensure the different frequencies stayed in the range of a plucked guitar. The resolution is 2.5 Hz, and we plucked the guitar string at half its length at 0.3275 m. Playing at that location: we only have odd harmonic. Once we measured through \textit{Audacity} the frequencies, we calculated the relative error compared to the analytical solution calculated previously. 

\begin{figure}[H]
    \centering
    \includegraphics[width=0.7\textwidth]{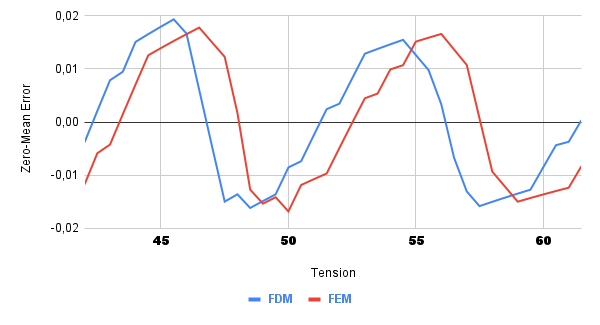}
    \caption{Zero-mean error between the measured and analytical fundamental frequency according to tension \(N\)}
    \label{fig:tensionmeanfonda}
\end{figure}

\begin{figure}[H]
    \centering
    \includegraphics[width=0.7\textwidth]{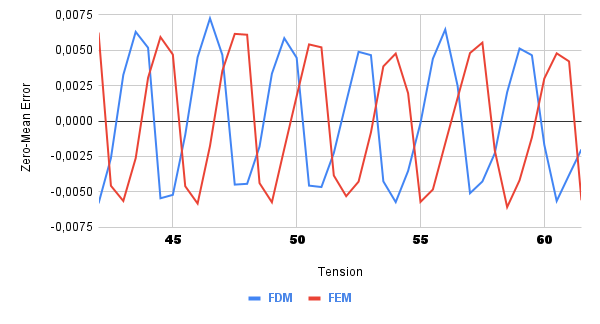}
    \caption{Zero-mean error of third harmonic frequency according to tension \(N\)}
    \label{fig:tensionmean3}
\end{figure}

\begin{figure}[H]
    \centering
    \includegraphics[width=0.7\textwidth]{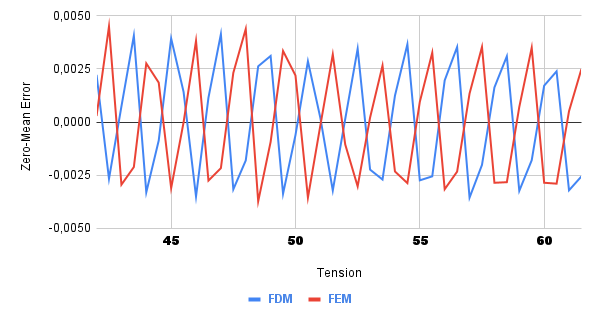}
    \caption{Zero-mean error of fifth harmonic frequency according to tension \(N\)}
    \label{fig:tensionmean5}
\end{figure}

\begin{figure}[H]
    \centering
    \includegraphics[width=0.7\textwidth]{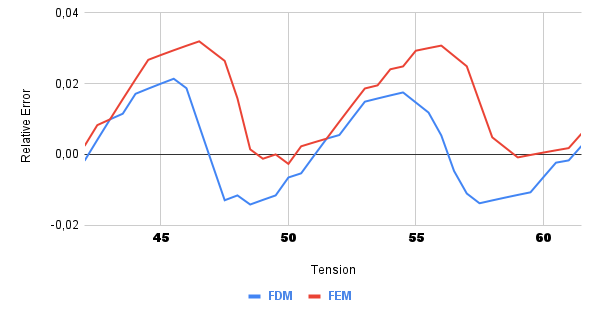}
    \caption{Error of fundamental frequency according to tension}
    \label{fig:tensionfundaerror}
\end{figure}

\begin{figure}[H]
    \centering
    \includegraphics[width=0.7\textwidth]{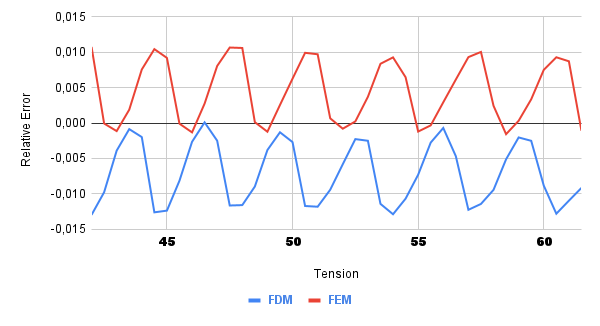}
    \caption{Error of third harmonic frequency according to tension}
    \label{fig:tension3error}
\end{figure}

\begin{figure}[H]
    \centering
    \includegraphics[width=0.7\textwidth]{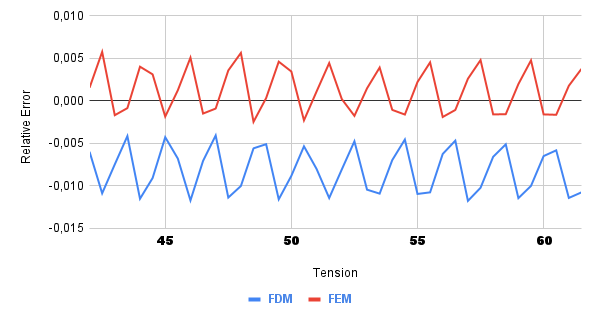}
        \caption{Error of fifth harmonic frequency according to tension}
    \label{fig:tension5error}
\end{figure}

The zero-mean error centered around zero (Fig.\ref{fig:tensionmeanfonda}, Fig.\ref{fig:tensionmean3} and Fig.\ref{fig:tensionmean5}) was then plotted. A periodicity of the error is visible for both methods. Both models follow the same error variation with a phase shift as if they complemented each other. The periodicity of this error is very particular. The plots with the means are visible in Figures \ref{fig:tensionfundaerror}, \ref{fig:tension3error}, and \ref{fig:tension5error}. 


In Figure \ref{fig:tensionmeanfonda}, the fundamental frequency $f_1$ has a tension periodicity of 9.5 N corresponding to 25 Hz. Examining the higher harmonics (Fig.\ref{fig:tensionmean3}, Fig.\ref{fig:tensionmean5}) we get:
\begin{itemize}
    \item $3f_1$: Periodicity 3.10 N or 8 Hz.
    \item $5f_1$: Periodicity 1.9 N or 5 Hz.
\end{itemize}
By computing the periodicity ratios:

\begin{equation}
    \frac{9.5}{3} = 3.17, \quad \frac{9.5}{5} = 1.9
\end{equation}

\begin{equation}
    \frac{25}{3} = 8.3, \quad \frac{25}{5} = 5
\end{equation}
We observe a clear relationship between the harmonic order and the periodicity of the error. The periodicity of the error is the harmonic order times the periodicity of error of the fundamental.

These periodicity errors may arise due to the discretization in time. Since we sample the frequency domain at discrete intervals, numerical artifacts manifest at harmonic positions. 
This behavior suggests that the discretization process influences the periodic detection of frequency variations in a structured manner. Moreover, the error margin is probably not only created by the simulation but also by the FFT that discretizes the signal in the frequency domain. The way the FFT is done impacts the result, but since the same FFT was done on each waveform, the tendency of the error of the simulation to be periodic should follow the same trend.


\subsection{Variation of the time interval}
In numerical simulations, the choice of time step precision significantly impacts the accuracy of results. During one second, we decreased the time interval and analyse how that affected the error. The parameters are fixed to:
\begin{itemize}
    \item Tension: \( 42.86 \) N
    \item Length: \( 0.655 \) \si{\meter}
    \item Position of plucking: \(0.18 \si{\meter}\)
    \item Number of nodes the string is split into: \( 80 \)
    \item Linear mass density: \( 4.30 \times 10^{-4} \) \si{\kilogram\per\meter}
\end{itemize}
A comparison of time-step values reveals that both methods have a minimum time-step to converge.
\begin{itemize}
    \item FDM converges when the time-step is  below $2.27 \times 10^{-5}$ \si{\second} 
    \item FEM converges when the time-step is  below $1.52 \times 10^{-5}$ \si{\second}
\end{itemize}
This discrepancy suggests that FEM requires a higher level of time precision than FDM to achieve comparable accuracy. The divergence of FEM at lower precision indicates its sensitivity to time discretization, emphasizing the need for finer resolution in FEM-based simulations\label{txt:nonNan}. For non-NaN values, the data step of \( 6.7 \times 10^{-4} \) was repeated 40 times, and the string was plucked at a location of \( 0.18 \) \si{\meter}. The plucking position was changed from the previous experience to see if the periodicity was due to the location. We thus have all the odds and even harmonics. Moreover, only the first three frequencies were measured. 

\begin{figure}[H]
    \centering
    \includegraphics[width=1\textwidth]{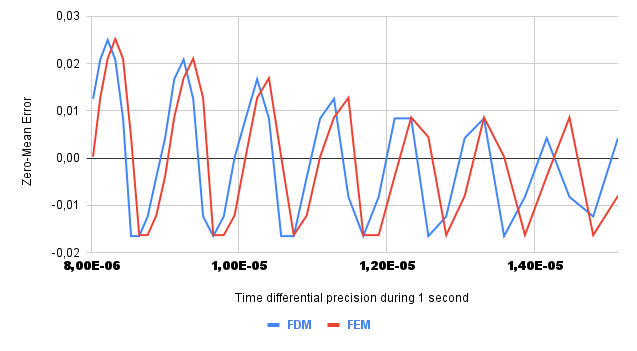}
    \caption{Zero-mean error of the fundamental frequency compared to analytical when changing the time step}
    \label{fig:timefundamean}
\end{figure}
\begin{figure}[H]
    \centering
    \includegraphics[width=1\textwidth]{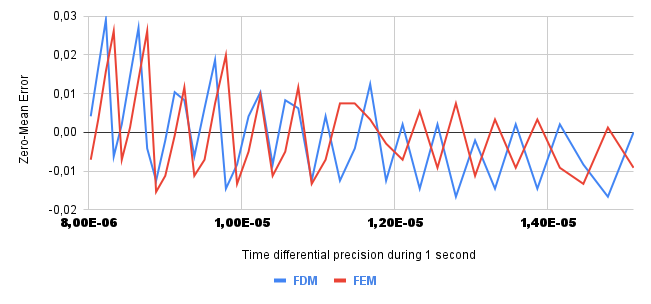}
    \caption{Zero-mean error of the second harmonic compared to analytical when changing the time step}
    \label{fig:time2mean}
\end{figure}
\begin{figure}[H]
    \centering
    \includegraphics[width=1\textwidth]{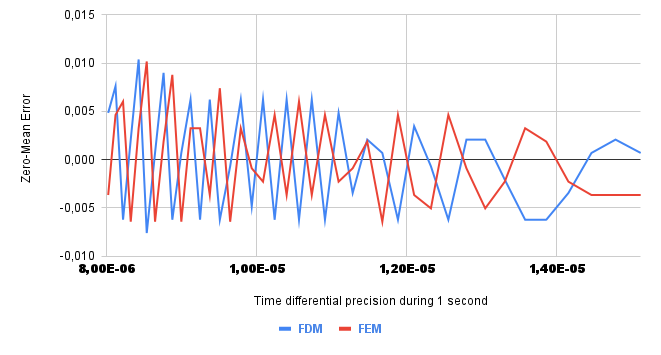}
    \caption{Zero-mean error of the third harmonic compared to analytical when changing the time step}
    \label{fig:time3mean}
\end{figure}
\begin{figure}[H]
    \centering
    \includegraphics[width=1\linewidth]{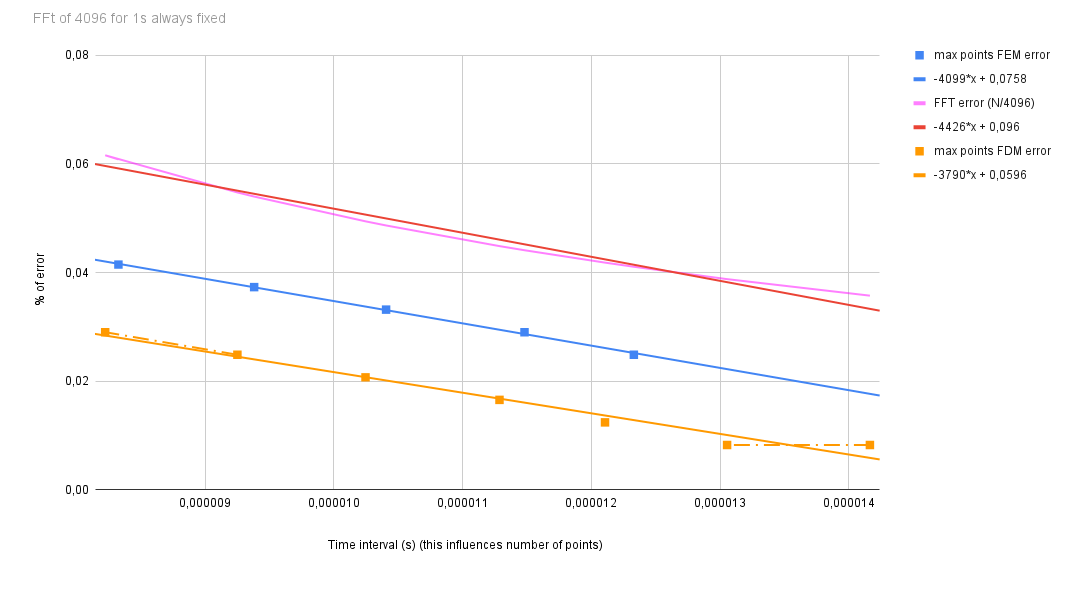}
    \caption{Fundamental error FFT frequency interval compared to Amplitude of errors when changing the time step}
    \label{fig:FFTerror}
\end{figure}
In the zero-mean error frequency plots (Fig.\ref{fig:timefundamean}, Fig.\ref{fig:time2mean}, Fig.\ref{fig:time3mean}), once again, it can be seen that both FEM and FDM exhibit the same evolution in error precision. However, the smaller the time interval, the more points we have in time, and the larger the error becomes. On the first array, the data does not seem to follow the expected curve of decreasing error with increased precision of the time measurement. This phenomenon is actually due to the way the FFT is done. The window size (i.e. number of points) remained the same even as the sampling frequency, directly linked to the time interval, increased. As a result, the error in the FFT measurement also increased. Figure \ref{fig:FFTerror} represents the evolution of the fundamentals' error precision as the sampling frequency increases. The curves represent the evolution of the fundamentals' error precision due to the FFT, the measurement in the FEM, and the measurement in the FDM. All three curves follow the same trend. This explains the previous results. The FFT influences the peak of the error; with fixed window size, the error max increases as the interval decreases. Finally, in the FFT, it is necessary to use a larger window size when increasing the sampling frequency, corresponding to reducing the time interval to reduce the error and not increase its peak.

\begin{figure}[H]
    \centering
    \includegraphics[width=1\textwidth]{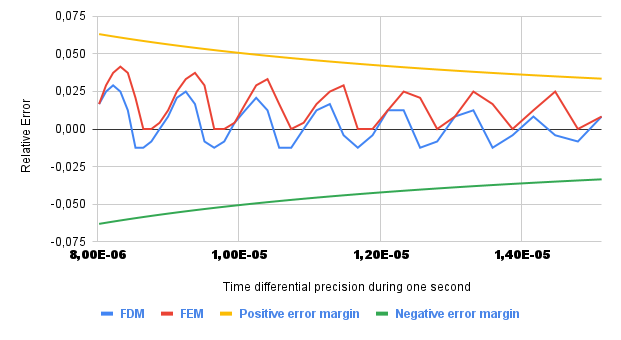}
    \caption{Error of the fundamental frequency compared to analytical when changing the time step}
    \label{fig:timefunda}
    \end{figure}

\begin{figure}[H]
    \centering
    \includegraphics[width=1\textwidth]{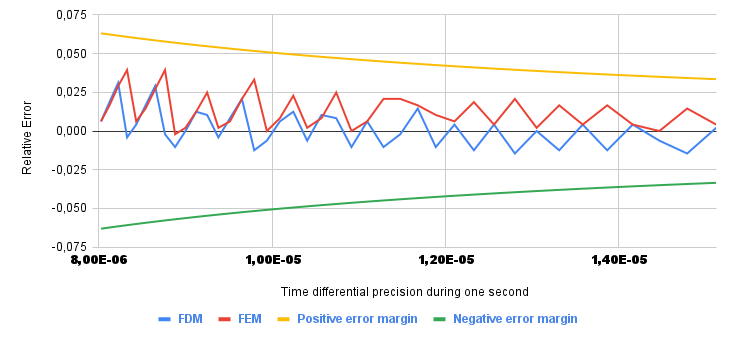}
    \caption{Error of the second harmonic compared to analytical when changing the time step}
    \label{fig:time2}
\end{figure}

\begin{figure}[H]
    \centering
    \includegraphics[width=1\linewidth]{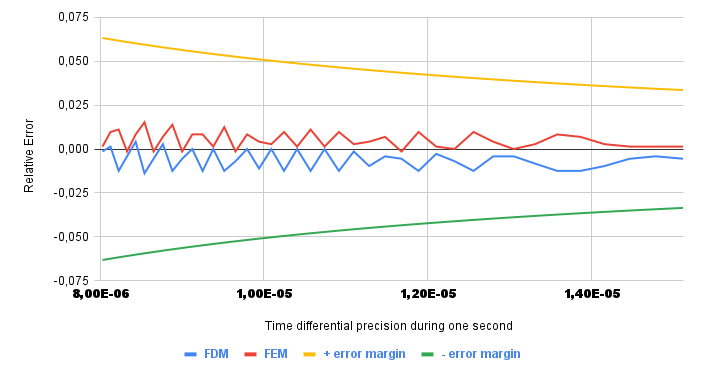}
    \caption{Error of the third harmonic compared to analytical when changing the time step}
    \label{fig:time3}
\end{figure}

Moreover, comparing in a broader perspective, figures \ref{fig:timefunda}, \ref{fig:time2}, and \ref{fig:time3} show the error with its mean. The error margin of the FFT is still visible. In the plot, both FEM and FDM follow the same error variations at the same rate. However, FDM converges for bigger time intervals than FEM (as explained in part \ref{txt:nonNan}), and the error max is smaller for FDM for the fundamental and second harmonic. 

\subsection{Variation in the number of nodes}
When doing FEM or FDM, you have several nodes you can create; for FEM, you can have a minimum of 1 element (2 nodes), and for FDM, you can also have at least 2 nodes. However, one element or two nodes would not create any sound in our case since both extreme nodes are fixed. Thus, we can cut the guitar string from three nodes to as many as we wish. Here, we see how the number of nodes affects the error. 
The fixed parameters are:
\begin{itemize}
    \item Tension: \( 42.86 \) N
    \item Length: \( 0.655 \) m
    \item Time Interval: \( 1.0 \times 10^{-5} \) \si{\second}
    \item Position of plucked: \(0.18 \si{\meter}\)
    \item Number of points in time: \( 1.0 \times 10^{5} \)
    \item Time of wave simulated: \( 1.0 \) \si{\second}
    \item Amplitude: \( 3.0 \times 10^{-4} \) \si{\meter}
    \item Linear mass density: \( 4.30 \times 10^{-4} \) \si{\kilogram\per\meter}
\end{itemize}
The guitar string is plucked, and all the harmonics are present.

\begin{figure}[H]
    \centering
    \includegraphics[width=1\linewidth]{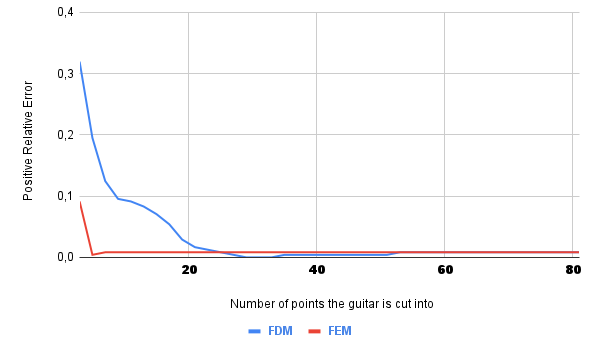}
    \caption{Positive relative error of the fundamental frequency compared to analytical according to the number of nodes}
    \label{fig:nnodefund}
\end{figure}
\begin{figure}[H]
    \centering
    \includegraphics[width=1\linewidth]{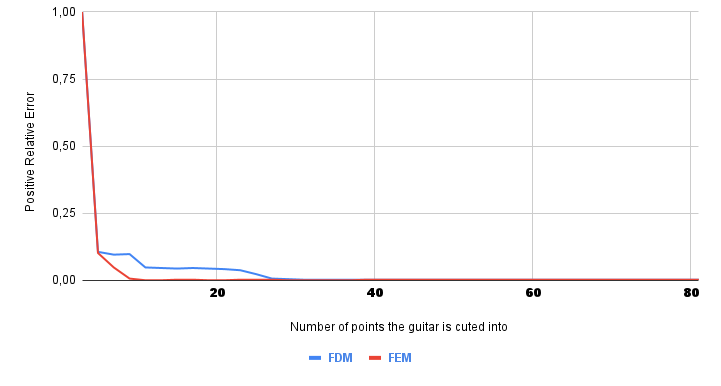}
    \caption{Positive relative error of the second harmonic compared to analytical according to the number of nodes}
    \label{fig:nnode2}
\end{figure}
\begin{figure}[H]
    \centering
    \includegraphics[width=1\linewidth]{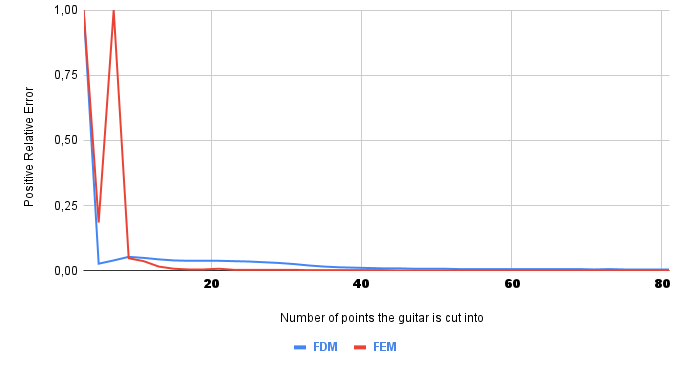}
    \caption{Positive relative error of the third harmonic compared to analytical according to the number of nodes}
    \label{fig:nnode3}
\end{figure}
\begin{figure}[H]
    \centering
    \includegraphics[width=1\linewidth]{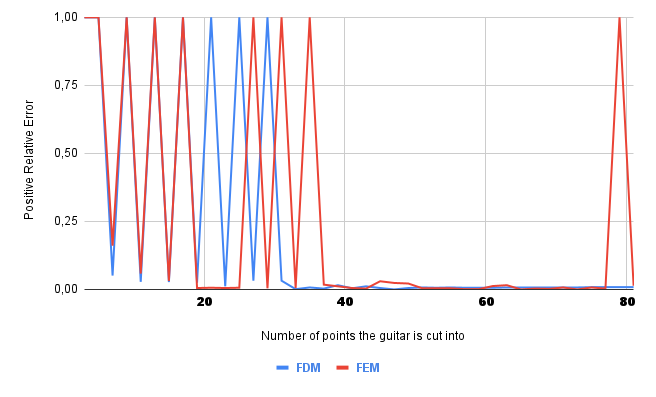}
    \caption{Positive relative error of the fourth harmonic compared to analytical according to the number of nodes}
    \label{fig:nnode4}
\end{figure}
\begin{figure}[H]
    \centering
    \includegraphics[width=1\linewidth]{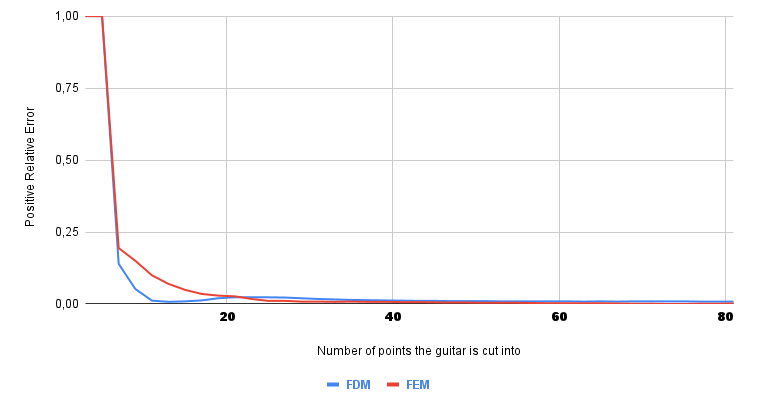}
    \caption{Positive relative error of the fifth harmonic compared to analytical according to the number of nodes}
    \label{fig:nnode5}
\end{figure}

Figures \ref{fig:nnodefund}, \ref{fig:nnode2}, \ref{fig:nnode3}, \ref{fig:nnode4} and \ref{fig:nnode5}, show the relative error of the measured frequency according to the number of nodes we cut the string into. When the error is equal to one (Fig.\ref{fig:nnode4}), no frequency is detected. The fundamental FEM converges faster than FDM for the fundamental (Fig.\ref{fig:nnodefund}) and the second harmonic (Fig.\ref{fig:nnode2}). However, after 40 nodes, the error is equivalent, and they both converge to the same error for most harmonics (Fig.\ref{fig:nnodefund}, Fig.\ref{fig:nnode2}, Fig.\ref{fig:nnode3} and Fig.\ref{fig:nnode5}). The fourth harmonic (Fig.\ref{fig:nnode4}) was hard to detect because it had a small amplitude, but some convergence was still detectable after 40 nodes were simulated. Finally, FDM converges faster in terms of the number of nodes representing an object than FEM.
\subsection{Analyzing the effects of modified parameters on the time of simulation}
\begin{figure}[H]
    \centering
    \includegraphics[width=1\linewidth]{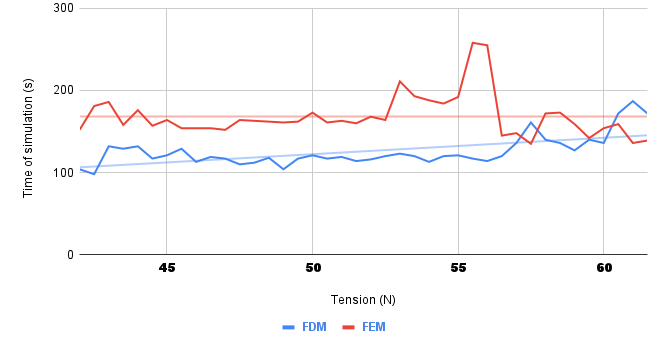}
    \caption{Time of FEM and FDM simulations of a string according to tension}
    \label{fig:timetension}
\end{figure}

\begin{figure}[H]
    \centering
    \includegraphics[width=1\linewidth]{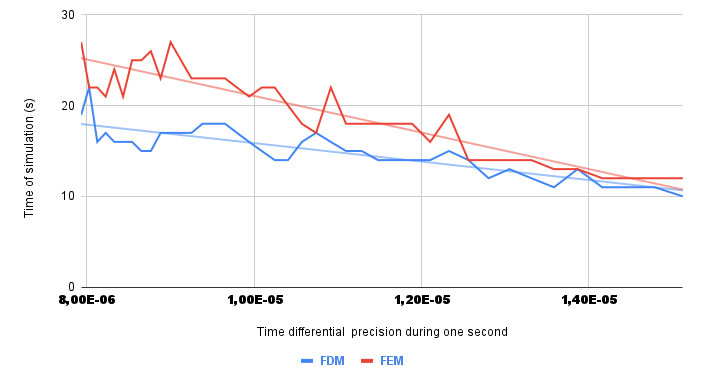}
    \caption{Time of FEM and FDM simulations according to the time interval chosen in the discretizations}
    \label{fig:timetime}
\end{figure}

\begin{figure}[H]
    \centering
    \includegraphics[width=1\linewidth]{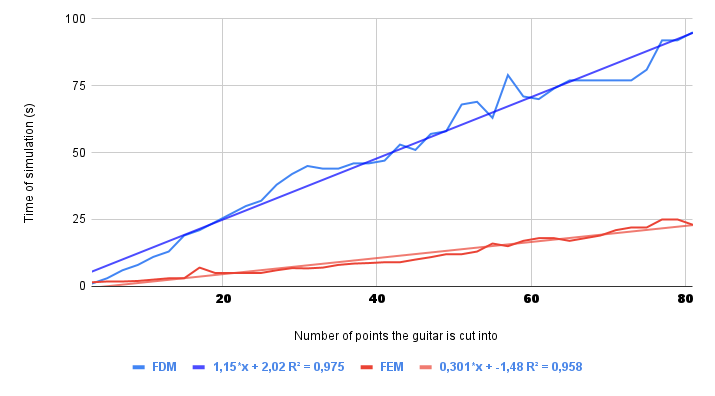}
    \caption{Time of FEM and FDM simulations of a string according to the number of cuts}
    \label{fig:timecuts}
\end{figure}

Two simulation codes, based on the FEM and the FDM, were developed in Python. Both codes were executed on the same computer, and the simulation times were recorded for each run corresponding to the experimental cases previously described. It should be emphasized that these performance results (Fig.\ref{fig:timetension}, Fig.\ref{fig:timetime}, Fig.\ref{fig:timecuts}) are inherently dependent on the specific computing environment as well as the implementation details of the Python code. Consequently, the findings presented here should not be considered universally representative but rather indicative within the context of the conditions under which the analysis was conducted.

Figure \ref{fig:timetension} underlines how an evolution in the tension affects the simulation time. The FEM curve is approximately following a horizontal line. Thus, the tension (Fig.\ref{fig:timetension}), so the frequency, does not influence the time. However, we can see that the FDM line has a small positive slope compared to FEM. However, FDM is still faster than FEM in these conditions. Following this trend, FDM would be slower than FEM after a certain frequency.  

Figure \ref{fig:timetime} illustrates the influence of a change of the time interval on the time of a simulation. Nevertheless, the smaller the time interval the slower the simulation since it means there are more data to calculate. FDM, compared to FEM, is quicker for the given time interval.

Figure \ref{fig:timecuts} illustrates the impact of varying the number of nodes composing the string on simulation time for both the FEM and the FDM. As the number of nodes increases, FEM demonstrates significantly shorter simulation times compared to FDM. This discrepancy can be attributed to differences in code implementation: the FDM approach relies on an explicit Python loop that iterates over all nodes, while the FEM implementation utilizes matrix operations applied directly to the entire set of elements. As a result, FEM is expected to offer greater computational efficiency. It is important to note that this analysis reflects the specific coding strategies employed and may vary under different implementations.
\section{To compare to reality: guitar with damping and without the modulus of elasticity}
The D'Alembert equation for transverse vibrations of a string with fluid damping is given by \cite{ajp1930} \cite{derveaux2002}:
\begin{equation}
\mu\frac{\partial^2 u}{\partial t^2} - T \frac{\partial^2 u}{\partial x^2} + \sigma \frac{\partial u}{\partial t} = 0,
\end{equation}
where:
\begin{itemize}
    \item \(u(x,t)\): Transverse displacement of the string at position \(x\) and time \(t\)
    \item \(c = \sqrt{\frac{T}{\mu}}\): Wave velocity in \si{\meter\per\second}, dependent on the tension \(T\) in \(N\) and the linear mass density \(\mu\) in \si{\kilogram\per\meter}
    \item \(\sigma\): Fluid damping coefficient \si{\kilogram\per\second}
\end{itemize}
This equation includes a damping term \(\sigma \frac{\partial u}{\partial t}\), which represents the resistance due to friction. We discretize the string in space and time:
\begin{itemize}
    \item The spatial domain \([0, L]\) is divided into \(N\) nodes with spacing \(\Delta x = \frac{L}{N}\).
    \item The temporal domain \([0, \tau]\) is divided into \(K\) points with spacing \(\Delta t = \frac{\tau}{K}\).
\end{itemize}
Using finite differences, we approximate:
\begin{align}
\frac{\partial^2 u}{\partial x^2} &\approx \frac{u_{n+1,k} - 2u_{n,k} + u_{n-1,k}}{\Delta x^2}, \\
\frac{\partial^2 u}{\partial t^2} &\approx \frac{u_{n,k+1} - 2u_{n,k} + u_{n,k-1}}{\Delta t^2}, \\
\frac{\partial u}{\partial t} &\approx \frac{u_{n,k+1} - u_{n,k-1}}{2 \Delta t}.
\end{align}
Substituting these into the D'Alembert equation yields:
\begin{equation}
0 = \mu \frac{u_{n,k+1} - 2u_{n,k} + u_{n,k-1}}{\Delta t^2} - T \frac{u_{n+1,k} - 2u_{n,k} + u_{n-1,k}}{\Delta x^2} + \sigma \frac{u_{n,k+1} - u_{n,k-1}}{2 \Delta t}
\end{equation}
Rearranging terms, and starting with the equation:
\begin{align}
    u_{n,k+1} \left( \mu \frac{1}{\Delta t^2} + \frac{\sigma}{2\Delta t} \right) 
    &= \frac{T}{\Delta x^2} \left( u_{n+1,k} + u_{n-1,k} \right) 
    \nonumber \\
    &+ \left( 2 \mu \frac{1}{\Delta t^2} - 2 \frac{T}{\Delta x^2} \right) u_{n,k} \quad \nonumber \\
    &+ u_{n,k-1} \left( - \mu \frac{1}{\Delta t^2} + \frac{\sigma}{2 \Delta t} \right)
\end{align}
To isolate \( u_{n,k+1} \), divide through by \( \mu \frac{1}{\Delta t^2} + \frac{\sigma}{2\Delta t} \):
\begin{equation}
    u_{n,k+1} = \frac{\frac{T}{\Delta x^2} \left( u_{n+1,k} + u_{n-1,k} \right) + \left( 2 \mu \frac{1}{\Delta t^2} - 2 \frac{T}{\Delta x^2} \right) u_{n,k} + u_{n,k-1} \left( - \mu \frac{1}{\Delta t^2} + \frac{\sigma}{2 \Delta t} \right)}{\mu \frac{1}{\Delta t^2} + \frac{\sigma}{2\Delta t}}
\end{equation}
We can define the coefficients to have a easier notation:
\begin{align}
\gamma &= \frac{T^2 \Delta t^2}{\Delta x^2}, \\
\alpha &= \mu + \frac{\sigma \Delta t}{2}, \\
\theta &= -\mu + \frac{\sigma \Delta t}{2}.
\end{align}
The equation becomes:
\begin{align}
u_{n,k+1} = \frac{\gamma \frac{1}{\Delta t^2} \left( u_{n+1,k} + u_{n-1,k} \right) + \left( 2 \frac{\mu}{\Delta t^2} - 2 \gamma \frac{1}{\Delta t^2} \right) u_{n,k} + u_{n,k-1} \left( \theta \frac{1}{\Delta t^2} \right)}{\alpha \frac{1}{\Delta t^2}}.
\end{align}
Thus, the final equation is:

\begin{equation}
u_{n,k+1} = \frac{\gamma}{\alpha} \left(u_{n-1,k} + u_{n+1,k}\right) + \frac{2(\mu - \gamma)}{\alpha} u_{n,k} + \frac{\theta}{\alpha} u_{n,k-1}
\end{equation}
The boundary conditions are the same as in the case without sampling seen in equation \ref{eq:bondstring}. The initial conditions are also the same as previously in equations \ref{eq:inistring1} and \ref{eq:inistring2}.

\subsection{Comparison to Reality}
\begin{figure}[H]
    \centering
    \includegraphics[width=0.5\linewidth]{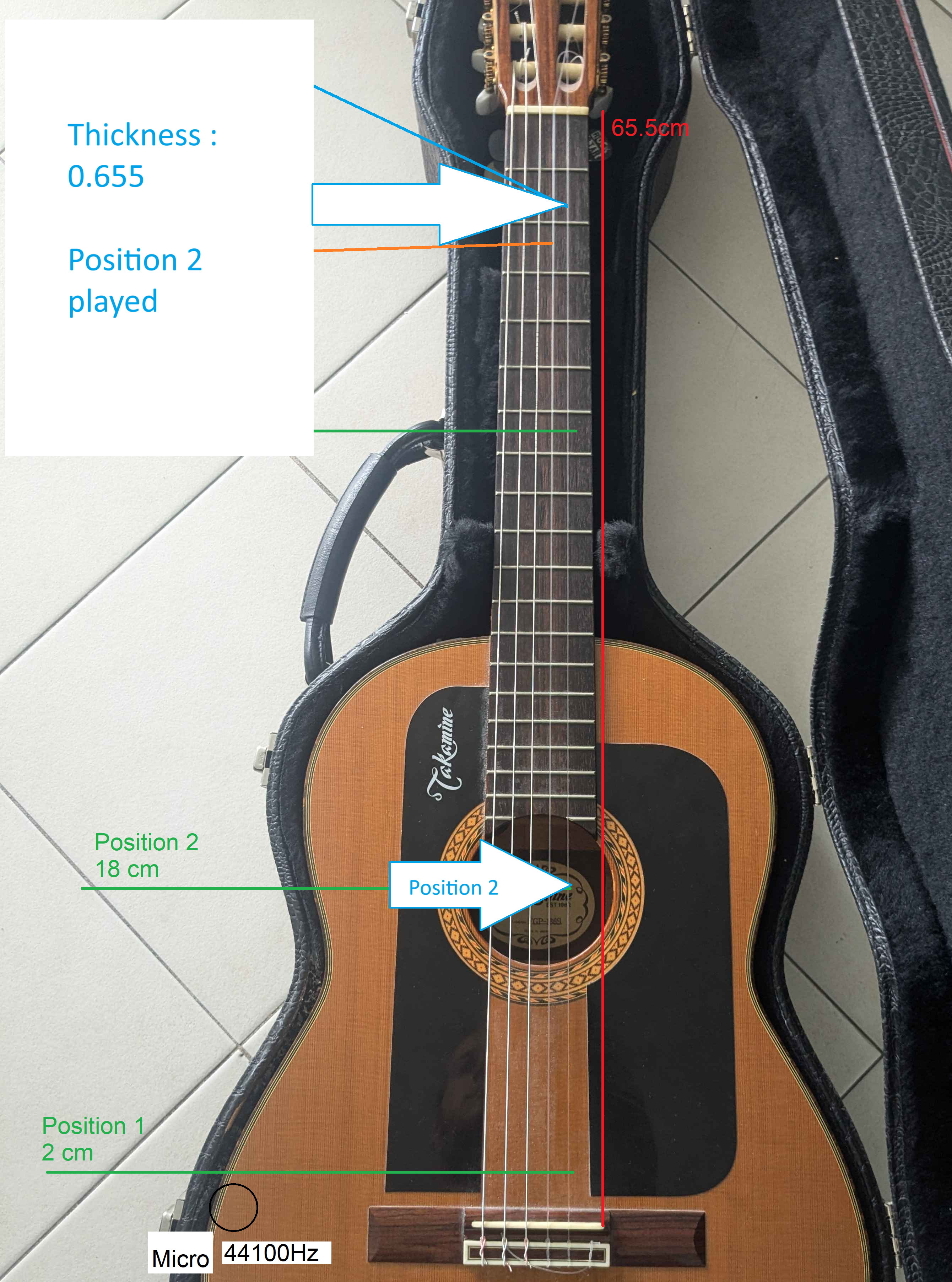}
    \caption{Real guitar of string length 0.655 \si{\meter} played on the fifth string at 0.18 \si{\meter}}
    \label{fig:realgui}
\end{figure}

This study compares our FDM simulation results to real-world conditions (Fig.\ref{fig:realgui}). The following physical parameters characterize the experimental setup, which we will use as the basis for our simulations:
\begin{itemize}
    \item Fundamental frequency: \( f = 246.9 \)\si{\hertz}
    (corresponding to the musical note B3), determined using a FFT with a window size of 4096.
    \item String dimensions:
    \begin{itemize}
        \item Length: 
        \(
        L = 65.5 \text{ cm}= 0.655 \si{\meter}
        \)
        \item Diameter: 
        \(
        d = 0.69 \text{ mm} = 6.9 \times 10^{-4} \si{\meter}
        \)
    \end{itemize}

    \item Linear mass density:
    \(
    \mu = 1150 \times \left(\frac{d}{2}\right)^2 \pi = 43 \times 10^{-5} \si{\kilogram\per\meter} = 4.30 \times 10^{-4} \si{\kilogram\per\meter}
    \)

    \item Playing position: The string is plucked at \(0.180 \si{\meter}\).
    \item Theoretical tension: 
    \(T = 45.02 \text{ N}\)
\end{itemize}
These parameters will be the reference for validating our numerical simulations against experimental data (Fig.\ref{fig:realgui}) \cite{polisano2016} \cite{td_ondes_mecaniques2017}. 

\begin{figure}[H]
    \centering
    \includegraphics[width=1\linewidth]{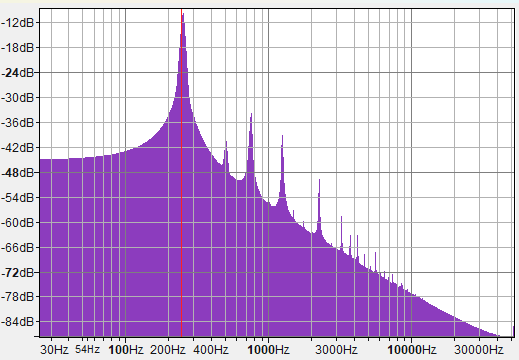}
    \caption{FFT through \textit{Audacity} on the simulated waveform, peak at 244 \(Hz\)}
    \label{fig:simsimfft}
\end{figure}

The simulation was performed under the following conditions:
\begin{itemize}
    \item Number of nodes: 
    \(
    80 \text{ nodes}
    \)
    \item Playing position: 
    \(
    22^{\text{nd}} \text{ point}
    \), 0.18 \si{\meter}
    \item Initial Amplitude: 
    \(
    3.0 \times 10^{-4} \si{\meter}
    \)
    \item Simulation duration: 
    3 \si{\second}
    \item Time step: 
    \(
    \Delta t = 9.65 \times 10^{-6} \si{\second}
    \)
    \item Damping
    \(
    \sigma = 0.0013 \si{\kilogram\per\second}
    \) (Chosen to match the time of damping of the real guitar)
\end{itemize}
To determine the appropriate values for our simulation for string time step (\(\Delta t\)) in the simulation, we analyzed the error introduced by the different parameters:
\begin{itemize}
    \item At \( T = 45.02 \) N, the frequency error is +2\% as shown in Figure \ref{fig:tensionfundaerror}.
    \item At \( \Delta t = 9.653 \times 10^{-6} \) \si{\second}, the frequency error is -1.2\% as shown in Figure \ref{fig:timefunda}.
\end{itemize}
Thus, knowing the error generated by FDM at that tension, a time differential is chosen to balance the error created by the FDM in the numerical simulation (Fig.\ref{fig:simsimfft}).
\begin{figure}[H]
    \centering
    \includegraphics[width=1\linewidth]{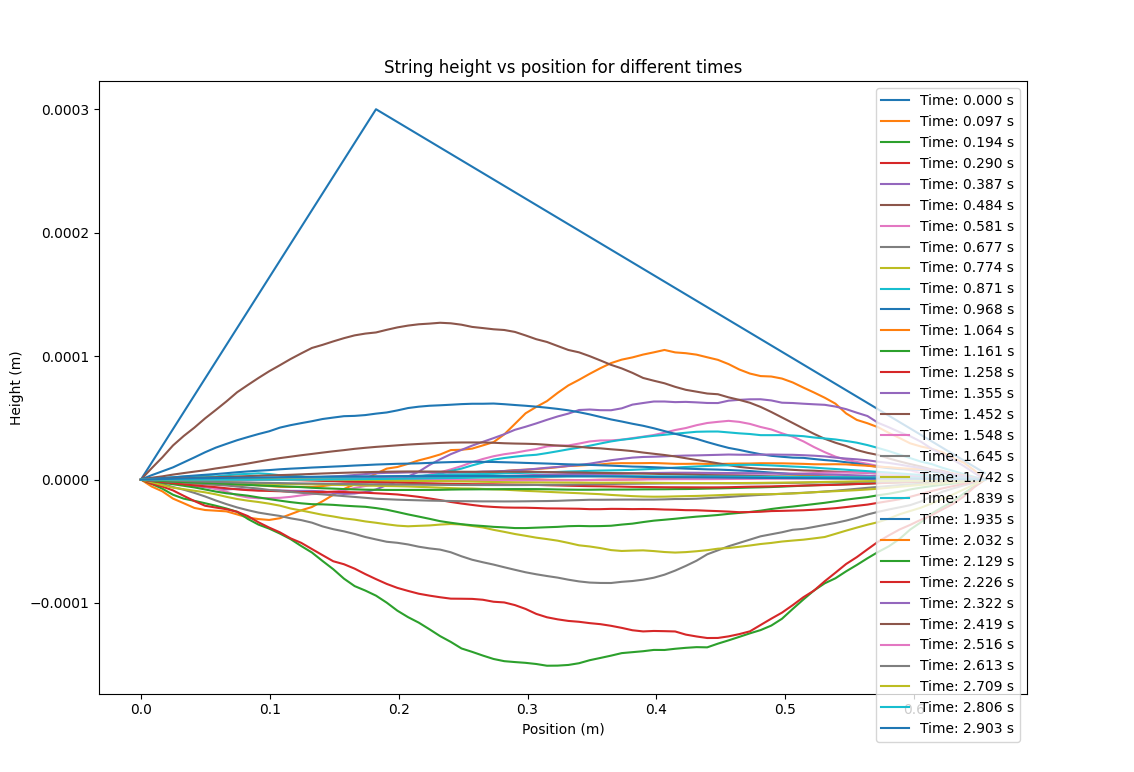}
    \caption{Simulation result of this particular string}
    \label{fig:simrea}
\end{figure}
\begin{figure}[H]
    \centering
    \includegraphics[width=1\linewidth]{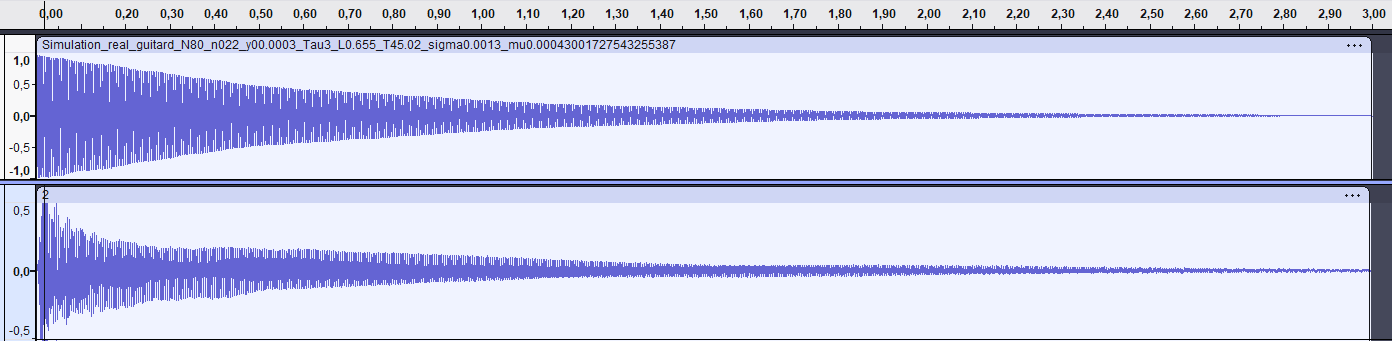}
    \caption{Damping looked at on \textit{Audacity}, the upper waveform is the simulation and the bottom one is the real signal}
    \label{fig:simreadamp}
\end{figure}
\begin{figure}[H]
    \centering
    \includegraphics[width=1\linewidth]{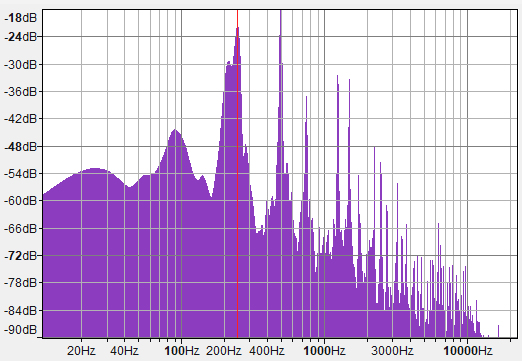}
    \caption{FFT through \textit{Audacity} of the real signal, peak at 252 \si{\hertz}}
    \label{fig:simreafft}
\end{figure}

Figure \ref{fig:simrea} shows the string simulated in time. Figure \ref{fig:simreadamp} shows the real and simulated signal damping. And finally, figure \ref{fig:simreafft}, and figure \ref{fig:simsimfft} show the FFT on both the real and the simulated waveform.
First, the damping is well represented, as shown in figure \ref{fig:simreadamp}. Then, in the frequency domain, the simulated signal (Fig.\ref{fig:simsimfft}) closely matches the real signal (Fig.\ref{fig:simreafft}), demonstrating the accuracy of the numerical model. However, it is essential to note that perceptually, the simulated sound still exhibits noticeable differences from the real instrument. Specifically, the simulation accurately represents the vibrating string (Fig.\ref{fig:simrea}). The same frequencies, are seen through the simulation and the experimental data. Finally, even if the spectrum has the same frequencies the simulation does not account for the reflections from surrounding surfaces or the effects of the guitar body, which significantly contribute to the instrument's characteristic timbre. The timbre is not represented in the simulation. However, if the goal is solely to identify the string type, the simulation can still achieve this. 

\subsection{Important note on the simulation}
When performing an FFT, it is crucial to clearly define the sampling rate and the sampling window, as errors can arise from these parameters. The accuracy of the FFT is inherently limited by the resolution of the discrete data and the chosen windowing function (section \ref{info:windowchoice}). Moreover, achieving higher numerical precision in the simulation does not necessarily translate to increased accuracy in practical measurements. In reality, experimental data acquisition is subject to various sources of uncertainty, including instrument limitations and environmental noise. Additionally, the FFT introduces a loss of precision due to spectral leakage and finite resolution. Therefore, while an exact numerical approach may be beneficial for theoretical analysis, it is essential to acknowledge the inherent limitations of real-world measurements.
\section{The cycling bell with damping and modulus of elasticity}
\subsection{Model}
\begin{figure}[H]
    \centering
    \includegraphics[width=0.5\linewidth]{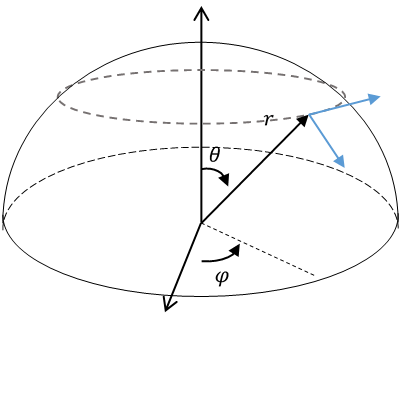}
    \caption{The model of cycling bell}
\end{figure}
The bell is modeled as the shell of a hemisphere with radius $R$, positioned relative to its center in spherical coordinates $(r, \theta, \varphi)$ as shown in the figure. We have a fixed point at the top. Assuming the shell of the bell is very thin (thickness $h \ll R$), for every point on the shell, it is possible to write: $r \approx R$.

\subsection{Theory}
As mentioned previously, we did not consider the flexion in the d'Alembert equation. It was a negligible term. However, that term is helpful for the cycling bell as it is mainly made of metal \cite{derveaux2022}. The fundamental equation of dynamics with damping with elasticity is expressed as follows:

\begin{equation}
    D\nabla^4 u + \sigma \frac{\partial u}{\partial t} + \rho h \frac{\partial^2 u}{\partial t^2} = 0
\end{equation}
where:
\begin{itemize}
    \item $D = \frac{E h^3}{12 (1 - \nu^2)}$: the bending rigidity in \si{\newton\meter}
    \item $h$: thickness in \si{\meter}
    \item $\rho$: density in \si{\kilogram\per\cubic\meter}
    \item $E$: Young's modulus in \si{\newton\per\square\meter}
    \item $\nu$: Poisson's ratio
    \item $\nabla^4$: the biharmonic operator (double Laplacian)
    \item $\sigma$: damping in \si{\newton\per\second\per\square\meter}
\end{itemize}
\subsection{Analytical resolution}
We seek solutions of the form:
\begin{equation}
    u(\theta, \varphi, t) = u(\theta, \varphi) e^{i \omega t}
\end{equation}
Substituting into the equation gives:
\begin{equation}
    \nabla^4 u - \left( \omega^2 \frac{\rho h}{D} - i \omega \frac{\sigma}{D} \right) u(\theta, \varphi) = 0
\end{equation}
Or equivalently:

\begin{equation}
    (\nabla^4 - k_\omega^4) u(\theta, \varphi) = 0
\end{equation}
where:
\begin{equation}
    k_\omega^4 = \omega^2 \frac{\rho h}{D} - i \omega \frac{\sigma}{D}
\end{equation}
We can write:
\begin{equation}
    (\nabla^4 - k_\omega^4) = (\nabla^2 + k_\omega^2)(\nabla^2 - k_\omega^2)
\end{equation}
The solution can be split into two components:
\begin{equation}
    u(\theta, \varphi) = u_{+}(\theta, \varphi) + u_{-}(\theta, \varphi)
\end{equation}
With:
\begin{align}
    (\nabla^2 + k_\omega^2) u_{+}(\theta, \varphi) = 0
    \\ 
    (\nabla^2 - k_\omega^2) u_{-}(\theta, \varphi) = 0
\end{align}
In the spherical coordinate system, the Laplacian is expressed as:
\begin{equation}
    \nabla^2 u = \Delta u = \frac{1}{r^2} \frac{\partial}{\partial r} \left( r^2 \frac{\partial u}{\partial r} \right) + \frac{1}{r^2 \sin \theta} \frac{\partial}{\partial \theta} \left( \sin \theta \frac{\partial u}{\partial \theta} \right) + \frac{1}{r^2 (\sin \theta)^2} \frac{\partial^2 u}{\partial \varphi^2}
\end{equation}
Since the shell is very thin, we can assume that $r$ is constant and equal to $R$. Thus, the waves described by equation (22) propagate mainly along the bell's surface, with variations in $\theta$ and $\varphi$: $u = u(\theta, \varphi, t)$
\begin{align}
\frac 1{r^2\sin \theta }\frac{{\partial}}{{\partial}\theta }\left(\sin \theta \frac{{\partial}u}{{\partial}\theta
}\right)&=\frac 1{r^2\sin \theta }\left(\cos \theta \frac{{\partial}u}{{\partial}\theta }+\sin \theta
\frac{{\partial}^2u}{{\partial}\theta ^2}\right)
\\ &=\frac 1{r^2}\left(\cot \theta \frac{{\partial}u}{{\partial}\theta
}+\frac{{\partial}^2u}{{\partial}\theta ^2}\right)
\end{align}
Having  $r=R=\mathit{constant}$, the Laplacian becomes:
\begin{equation}
{\nabla}^2u=\frac 1{R^2}\left(\cot \theta \frac{{\partial}u}{{\partial}\theta }+\frac{{\partial}^2u}{{\partial}\theta
^2}+\frac 1{\left(\sin \theta \right)^2}\frac{{\partial}^2u}{{\partial}\varphi ^2}\right)
\end{equation}
We seek the solutions by separating variables:
\begin{equation}
\begin{matrix}
u\left(\theta ,\varphi \right) = w\left(\theta \right) \cdot h\left(\varphi\right) \\
\end{matrix}
\end{equation}
We write:
\begin{equation}
\left(\nabla^2 + k_{\omega}^2\right) w\left(\theta\right) \cdot h\left(\varphi\right) = 0
\end{equation}
Taking:
\begin{equation}
\left( \sin \theta \cos \theta \frac{\partial w}{\partial \theta} + \left(\sin \theta\right)^2 \frac{\partial^2 w}{\partial \theta^2} \right) \frac{1}{w\left(\theta\right)} 
+ \left(\sin \theta\right)^2 R^2 k_{\omega}^2 
= -\frac{\partial^2 h}{\partial \varphi^2} \frac{1}{h\left(\varphi\right)}
\label{eq:simplefieddalembert}
\end{equation}
In the equation \ref{eq:simplefieddalembert}, the term on the right-hand side depends only on $\varphi$, and the term on the left-hand side depends only on $\theta$. This implies that each term must equal a constant $\lambda$. We thus obtain:
\begin{equation}
\left\{\begin{matrix}\left(\sin \theta \cos \theta {\partial}\frac{w_{+}}{{\partial}\theta }+\left(\sin \theta
\right)^2{\partial}^2\frac{w_-}{{\partial}\theta ^2}\right)\frac 1{w_+} + \left(\sin \theta
\right)^2R^2k_{\omega }^2=\lambda \\-{\partial}^2\frac{h_{+}}{{\partial}\varphi ^2}\frac 1{h_{+}}=\lambda \end{matrix}\right.
\end{equation}
We can rewrite it:
\begin{equation}
\left\{\begin{matrix}\left(\sin \theta \right)^2{\partial}^2\frac{w_{+}}{{\partial}\theta ^2}+\sin \theta \cos \theta
{\partial}\frac{w_{+}}{{\partial}\theta }+\left(\left(\sin \theta \right)^2R^2k_{\omega }^2-\lambda \right)\cdot
w_{+}=0\\\lambda h_{+}+{\partial}^2\frac{h_{+}}{{\partial}\varphi
^2}=0\end{matrix}\right.
\end{equation}
Or:
\begin{equation}
\left\{\begin{matrix}{\partial}^2\frac{w_{+}}{{\partial}\theta ^2}+\cot \theta {\partial}\frac{w_{+}}{{\partial}\theta
}+\left(R^2k_{\omega }^2-\frac{\lambda }{\left(\sin \theta \right)^2}\right)\cdot w_{+}\left(\theta \right)=0\\\lambda
h_{+}+{\partial}^2\frac{h_{+}}{{\partial}\varphi ^2}=0 \label{eq:hplus}
\end{matrix}\right.
\end{equation}
We would also get:
\begin{equation}
\left\{\begin{matrix}{\partial}^2\frac{w_{-}}{{\partial}\theta ^2}+\cot \theta {\partial}\frac{w_{-}}{{\partial}\theta
}+\left(-R^2k_{\omega }^2-\frac{\lambda }{\left(\sin \theta \right)^2}\right)\cdot w_{-}\left(\theta
\right)=0\\\lambda h_{-}+\frac{{\partial}^2h_{-}}{{\partial}\varphi ^2}=0\label{eq:hminus}\end{matrix}\right.
\end{equation}
Let us specify the ranges of the variables in question:
\begin{align}
\varphi &{\in}\left[0; \pi 2\right[ (+2k\pi)\\
\theta &{\in}\left[0;\frac{\pi } 2\right]
\label{eq:inftheta}
\end{align}
The solutions to the second equality in \ref{eq:hplus} and \ref{eq:hminus} are exponential functions with:

\begin{align} h_+(\varphi)&=h_{0+}e^{\pm i\sqrt{\lambda }\varphi }\\
h_-(\varphi)&=h_{0-}e^{\pm i\sqrt{\lambda }\varphi }
\end{align} 
From a physical standpoint, the functions $h_{+}$ and $h_{-}$ are $2\pi$ periodic. This means that $\sqrt{\lambda}$ must be an integer, thus, $\sqrt{\lambda }=m$$ \in \mathbf{N}$.
By setting: $\Omega^2=R^2k_{\omega}^2$ we can now solve the equation \ref{eq:legendre}  below (for both $w_{+}$ and $w_{-}$).
\begin{equation} \frac{{\partial}^2w\left(\theta \right)}{{\partial}\theta^2} + \cot \theta \frac{{\partial}w\left(\theta \right)}{{\partial}\theta} + \left(\pm \Omega^2 - \frac{\lambda}{(\sin \theta)^2}\right)w\left(\theta \right) = 0 
\label{eq:legendre}
\end{equation} 
We can change the variable to find solutions to the equation \ref{eq:legendre}.
By setting $x=\cos \theta$ with $\theta$ respecting its interval \ref{eq:inftheta}, we can show that the equation can be rewritten with $x \in [0; 1]$ as:
\begin{equation} \left(1-x^2\right)\frac{{\partial}^2w\left(x\right)}{{\partial}x^2} - 2x\frac{{\partial}w\left(x\right)}{{\partial}x} + \left(\pm \Omega^2 - \frac{m^2}{1-x^2}\right)w\left(x\right) = 0 
\label{eq:reallegendrer}
\end{equation} 
The equation \ref{eq:reallegendrer} is called the "associated Legendre equation". We can rewrite the equation  \ref{eq:reallegendrer} by seeking a solution, with $m$ an integer, in the form of:
\begin{equation}
w\left(x\right)=\left(1-x^2\right)^{\frac m 2}p\left(x\right)
\label{eq:relationwx}
\end{equation}
In this case, the equation is rewritten with $x \in [0; 1]$ as:
\begin{equation} \left(1-x^2\right)\frac{{\partial}^2p\left(x\right)}{{\partial}x^2} - 2x\left(m+1\right)\frac{{\partial}p\left(x\right)}{{\partial}x} + \left(\pm \Omega^2 - m\left(m+1\right)\right)p\left(x\right) = 0 
\label{eq:plegendre}
\end{equation} 
We seek the solution $p\left(x\right)$ in the form of a convergent power series on $x \in [0; 1]$:

\begin{align} p\left(x\right) &= \sum_{n \geq 0} a_n x^n
\label{eq:firstsub}
 \\ \frac{{\partial}p\left(x\right)}{{\partial}x} &= \sum_{n \geq 0} a_{n+1} \left(n+1\right) x^n 
 \label{eq:secondsub}
 \\
 \frac{{\partial}^2p\left(x\right)}{{\partial}x^2} &= \sum_{n \geq 0} a_{n+2} \left(n+2\right)\left(n+1\right) x^n 
 \label{eq:finalsub}
 \end{align} 
Substituting equation \ref{eq:finalsub}, \ref{eq:firstsub}, and \ref{eq:secondsub} into equation \ref{eq:plegendre}, we get:

\begin{align}
0 = \sum_{n \geq 0} a_{n+2} \left(n+2\right)\left(n+1\right) x^n &- \sum_{n \geq 0} a_{n+2} \left(n+2\right)\left(n+1\right) x^{n+2} 
\notag \\
- \sum_{n \geq 0} a_{n+1} \left(n+1\right) \left(m+1\right) x^{n+1} &+ \sum_{n \geq 0} a_n \left(\pm \Omega^2 - m\left(m+1\right)\right) x^n
\end{align} 
This simplifies to:

\begin{align} 
0 = \sum_{n \geq 0} a_{n+2} \left(n+2\right)\left(n+1\right) x^n - \sum_{n \geq 2} a_n \left(n\right)\left(n-1\right) x^n \notag \\
- \sum_{n \geq 1} a_n \left(n\right) \left(m+1\right) x^n + \sum_{n \geq 0} a_n \left(\pm \Omega^2 - m\left(m+1\right)\right) x^n  
\end{align} 
This gives us that for $n \geq 0$, the coefficients of $x^n$ are:

\begin{equation}
a_{n+2}\left(n+2\right)\left(n+1\right)-a_n\cdot \left(n\left(n+m\right)+m\left(m+1\right){\mp}\Omega ^2\right)
\end{equation}
By substituting the derivatives of $p(x)$ into the equation \ref{eq:plegendre} and by thinking through recurrence, we can prove that to have all the coefficients at zero, we must have:

\begin{equation} a_{n+2} = a_n \frac{n^2 + n \left(2m+1\right) + m \left(m+1\right) {\mp} \Omega^2}{\left(n+2\right)\left(n+1\right)} \end{equation} 
\subsubsection{First case}
To avoid divergence of the series (even when $x=1$), it is required that all the coefficients are zero from a certain rank onward.
This means that the suitable values of $\Omega$ must satisfy for some $n_0$:
\begin{equation} \Omega = \pm \sqrt{n_0^2 + n_0 \left(2m+1\right) + m \left(m+1\right)} 
\label{eq:omeganonecomp}
\end{equation} 
We can also see that there is a complex solution:

\begin{equation} \Omega = \pm i  \sqrt{n_0^2 + n_0 \left(2m+1\right) + m \left(m+1\right)} 
\label{eq:omegacomp}
\end{equation}
It should be noted that the recurrence relation between the coefficients of the power series is of order 2. This means that for all the coefficients to be zero from a certain rank onward, all coefficients must have the same parity as $n_0$.
By setting $k = n_0 + m$, in equation \ref{eq:omegacomp} and equation \ref{eq:omeganonecomp}, \(\Omega \) can be either:
\begin{align}
\Omega &=\pm \sqrt{k\left(k+1\right)}
\label{eq:omegareal}
\\
\Omega &=\pm i \sqrt{k\left(k+1\right)}
\label{eq:omeganotreal}
\end{align}
\subsubsection{Second case}
\begin{figure}[H]
    \centering
    \includegraphics[width=0.65\linewidth]{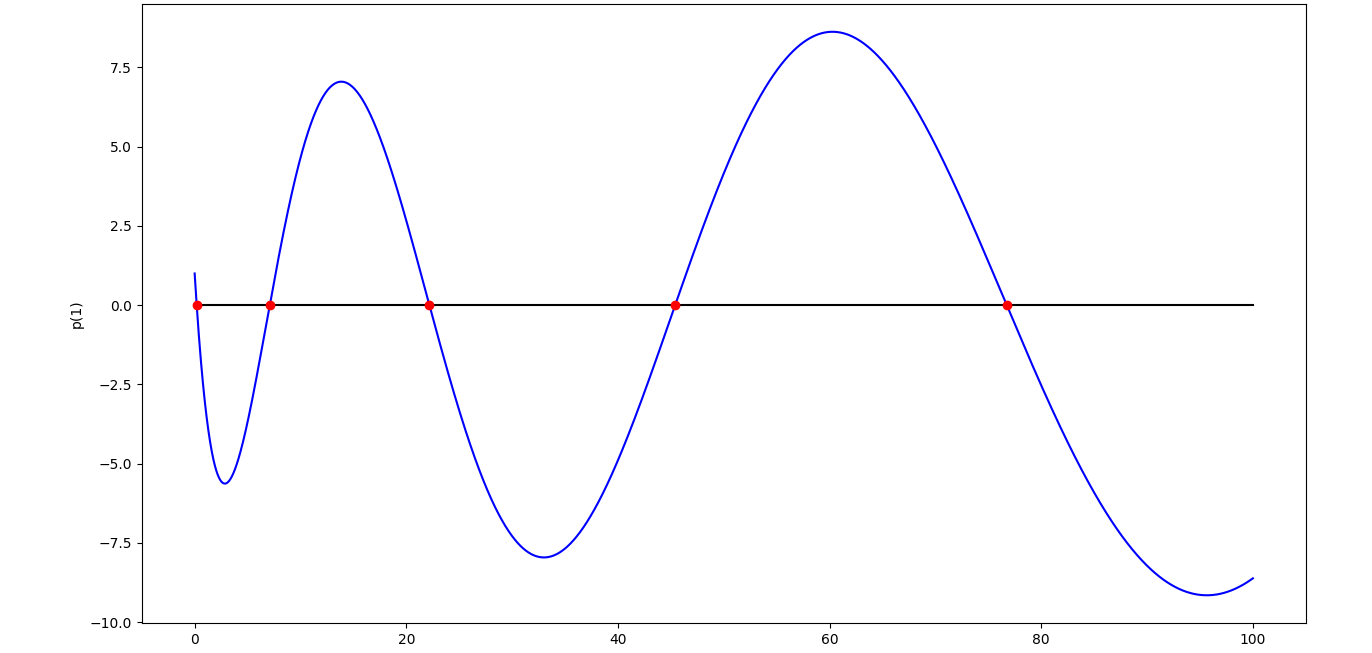}
    \caption{$p(1)$ according to $\Omega$ in \(m = 0\)}
    \label{fig:theobell}
\end{figure}

We seek convergence for all \(x \leq 1\).
D'Alembert's rule says convergence for all \(x < 1\).
But at \(x = 1\) we need to use:
Raabe-Duhamel test for \(x=1\)
\begin{equation}
    \left| \frac{a_2(k+1)}{a_2k} \right| = 1 - \frac{2 - m}{k} + o\left( \frac{1}{k} \right)
\end{equation}
Thus, to avoid divergence, we will have convergence if \(m = 0\); however, when that is the case, \( w(x) = p(x)\) (equation \ref{eq:relationwx}). With the condition at the limit, the top is fixed \(p(1) = 0\). Figure \ref{fig:theobell} shows the value of $p(1)$ according to different $\Omega$ for $m=0$. It highlights all the $\Omega$ where $p(1) = 0$. It has been verified that all the values of $\Omega$ where the plot passed by zero are the cases where there is a finite sum; thus, $\Omega$ follows the first case seen before where all the terms are null after a rank $n_0$. In conclusion, the only case where that condition is verified is when all the terms are null after a certain rank.

Taking the equations \ref{eq:omegareal} and \ref{eq:omeganotreal}, the frequency solutions are either way, because \(i^4 = 1\) so \(\Omega^4 = (Rk_{\omega })^4\):
\begin{equation} \omega^2 \frac{\mathit{\rho h} R^4}{D} - \mathit{i \omega} \frac{\sigma R^4}{D} - k^2 \left(k+1\right)^2 = 0 \end{equation} 
This means:
\begin{equation} \omega = \left(i \frac{\sigma}{2 \mathit{\rho h}} \pm \sqrt{- \left( \frac{\sigma}{2 \mathit{\rho h}} \right)^2 + \frac{D}{\mathit{\rho h} R^4} k^2 \left(k+1\right)^2} \right) \end{equation}
Finally we have an attenuation term $e^{\frac{-\sigma}{2 \mathit{\rho h}} t}$ and a harmonic term $e^{2i \mathit{\pi} f_k t}$ with:
\begin{equation} f_k = \frac{1}{2 \pi} \sqrt{- \left( \frac{\sigma}{2 \mathit{\rho h}} \right)^2 + \frac{D}{\mathit{\rho h} R^4} k^2 \left(k+1\right)^2} \end{equation}
 We calculate two characteristic coefficients of the material and the geometry of the shell:

\begin{itemize}
\item The attenuation coefficient $\gamma = \frac{\sigma}{2 \mathit{\rho h}}$
\item The stiffness coefficient $\alpha = \frac{D}{\mathit{\rho h} R^4}$ \end{itemize} 
Next, we choose two integers $m_0$ and $n_0$ that determine the vibration mode. For each pair $(m_0, n_0)$ chosen, we determine the coefficient $k_0 = n_0 + m_0$ which allows the calculation of the vibration frequency:

\begin{equation} f_{k_0} = \frac{1}{2 \pi} \sqrt{-\gamma^2 + \alpha k_0^2 \left(k_0 + 1\right)^2} \end{equation} 
The vibration mode is written as follows:
\begin{equation} u\left(\theta, \varphi, t\right) = u_{(m_0, n_0)}\left(\theta, \varphi\right) e^{\frac{-\sigma}{2 \mathit{\rho h}} t} e^{2i \pi f_{k_0} t} 
\label{eq:bellfinal}
\end{equation} 
The elongations are written for the chosen $m_0$ and $n_0$ as:
\begin{equation} u_{(m_0, n_0)} = w_{(m_0, n_0)}\left(\theta\right) \cdot h_{m_0}\left(\varphi\right) \end{equation} 
With:

\begin{align}
h_{m_0}\left(\varphi \right)&=h_0e^{im_0\varphi }
\\
w_{(m_0,n_0)}\left(\theta \right)&=\left(\sin \theta \right)^{\frac{m_0} 2}\cdot p_{n_0}\left(\cos \theta \right)
\\
p_{n_0}\left(x\right)&=a_0+a_1x^1+a_2x^2+\dots{}a_{n_0}x^{n_0}
\end{align}
To calculate the coefficients of $p_{n_0}\left(x\right)$, we first calculate:

\begin{equation} \Omega_0 = \sqrt{k_0 \left(k_0 + 1\right)} \end{equation} If $n_0$ is even, all coefficients $a_n$ with odd $n$ are zero. If $n_0$ is odd, all coefficients $a_n$ with even $n$ are zero. Next, for $n$ varying from 0 to $n_0$, we calculate:

\begin{equation} a_{n+2} = a_n \frac{n^2 + n \left(2 m_0 + 1\right) + m_0 \left(m_0 + 1\right) - \Omega_0^2}{\left(n+2\right)\left(n+1\right)} \end{equation} 
We start with $a_0$ for the even coefficients and with $a_1$ for the odd coefficients. Due to the recurrence relation between the coefficients, we notice that all even coefficients are proportional to $a_0$ and all odd coefficients are proportional to $a_1$.
Thus, the coefficients can be adjusted based on the boundary conditions.
Parameters for numerical application:
\begin{align*}
    \sigma &= 10  \si{\kilogram\per\square\meter} 
    \\ R &= 0.05 \si{\meter} 
    \\ h &= 0.001 \si{\meter} 
\end{align*}

\begin{table}[h!]
\centering
\caption{Material properties: density, Young’s modulus $E$, and Poisson’s ratio $\nu$}
\begin{tabular}{|l|c|c|c|}
\hline
Material & Density (\si{\kilogram\per\cubic\meter}) & $E$ (\si{\giga\pascal}) & $\nu$ \\
\hline
Steel & 7850 & 210 & 0.24–0.30 \\
Aluminum & 2700 & 62 & 0.24–0.33 \\
Copper & 8920 & 128 & 0.33 \\
Brass & 8470 & 80–100 & 0.37 \\
\hline
\end{tabular}

\end{table}

\subsubsection{Numerical test}
\begin{figure}[H]
    \centering
    \includegraphics[width=1\linewidth]{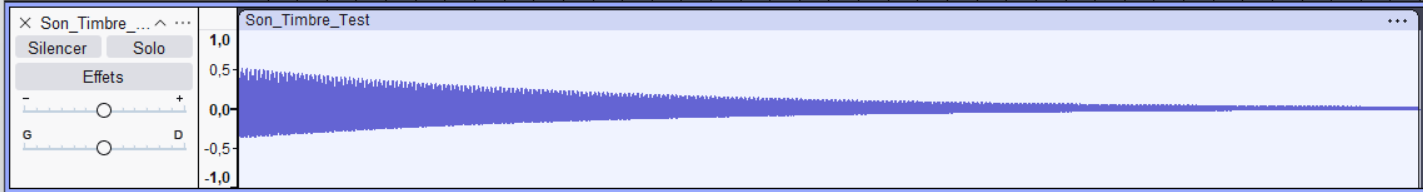}
    \caption{Time signal of the audio file generated viewed on \textit{Audacity}}
    \label{fig:resultwaveform}
\end{figure}

Using Matlab, we could compute the sound the bell would do in theory by superposing all the harmonics with the proper amplitudes (Fig.\ref{fig:resultwaveform}).

\subsection{Comparing analytical to reality }

\begin{figure}[H]
    \centering
    \includegraphics[width=0.8\linewidth]{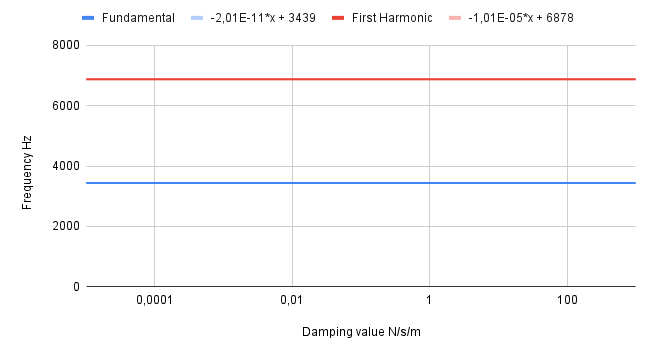}
    \caption{Damping effect on the fundamental frequency and first harmonic in logarithmic scale with linear regression}
    \label{fig:Dampingfft}
\end{figure}

\begin{figure}[H]
    \centering
    \includegraphics[width=0.25\linewidth]{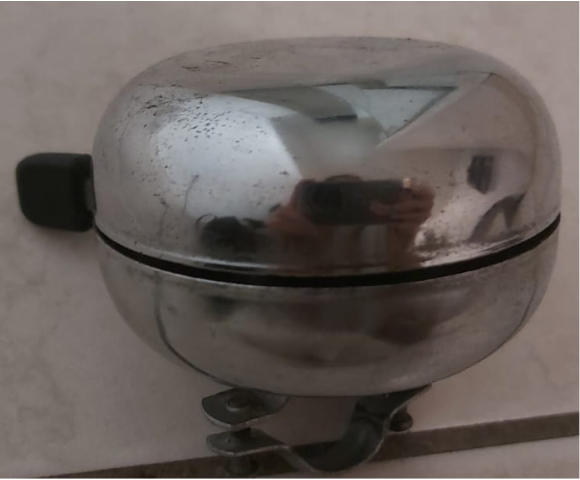}
    \caption{Aluminium bell}
    \label{fig:albell}
\end{figure}

\paragraph{Important information}
Before comparing our mathematical model to reality, we can see the numerical frequency is related to the damping term:
\begin{equation} f_k = \frac{1}{2 \pi} \sqrt{- \left( \frac{\sigma}{2 \mathit{\rho h}} \right)^2 + \frac{D}{\mathit{\rho h} R^4} k^2 \left(k+1\right)^2} 
\label{eq:frequencybell}\end{equation}
In the equation \ref{eq:frequencybell} showing the frequencies of a bell there is a damping term. Yet there are no known explicit damping terms for the cycling bell. Figure \ref{fig:Dampingfft} shows the effect of the term damping on the frequency in the logarithmic scale. The light orange and light blue lines coincide with the orange and blue lines. They represent linear regression, and both slop terms are negligible compared to the initialization terms. In consequence, the damping term has little effect on the frequency.
 Fixing the damping at $10 \si{\kilogram\square\meter\per\second}$ allows us to match the damping speed of a real-life cycling bell that was tested knowing that it will not effect the analysis comparing the experimental and the calculated frequencies. Figure \ref{fig:albell} shoes a big aluminum bell used in the experiment.

The parameters of the bell (Fig.\ref{fig:albell}) are:
\begin{itemize}
    \item Damping coefficient (fixed arbitrarily): 
    \(\sigma = 10  \si{\kilogram\per\square\meter\per\second}
    \)
    \item Density: 
    \( \rho = 2.70 \times 10^{3} \si{\kilogram\per\cubic\meter}\)
    \item Radius: 
    \( R = 4.00 \times 10^{-2} \si{\meter}\)
    \item Thickness: 
    \( h = 8.00 \times 10^{-4} \si{\meter}\)
    \item Young's modulus: 
    \( E = 6.20 \times 10^{10} \text{ N} \cdot \si{\per\square\meter}\)
    \item Poisson's ratio: \(
    \nu = 3.00 \times 10^{-1}\)
\end{itemize}

\begin{table}[H]
    \centering
    \caption{Comparison of fundamental and harmonic frequencies (\textbf{f$_n$}) between Matlab simulations, experimental data, and mathematical calculations for aluminum.}
    
    \begin{tabular}{|l|c|c|c|c|}
        \hline
        Cycling Bell & f$_1$ (Hz) & f$_2$ (Hz) & f$_3$ (Hz) & f$_4$ (Hz) \\
        \hline
        Model (Matlab) & 689 & 1385 & 2305 & 3462 \\
        \hline
        Reality & 688 & 1742 & 3080 & 3502 \\
        \hline
        Analytical analysis & 692 & 1385 & 2308 & 3462 \\
        
        \hline
    \end{tabular}
    \label{tab:frequency_comparison}
\end{table}
Table \ref{tab:frequency_comparison} shows the different frequencies calculated through different methods. The analytical analysis of the physical properties of the bell is one method. Another one is measuring frequencies through \textit{Audacity} of the experimental data. Finally, the frequencies were measured through \textit{Audacity} after simulating the analytical result through Matlab (summing all the vibration modes of equation \ref{eq:bellfinal}). The code of Matlab directly corresponds to the mathematical frequency; thus, it should give the same result for the frequencies. However, there are a few Hertz differences for the fundamental frequency and the third harmonic. Thus, the FFT applied to the Matlab simulation greatly influences the results.  The window is still at 4096; we are losing information and precision. Furthermore, the same FFT is applied to experimental data. In consequence, there is the same error margin. We also find almost the same frequency result as in the analytical case. 

Moreover, a few Hertz of difference is not detectable for an average human ear; this shows that the analytical result can fully be used to model the bell. The same experience was conducted with a supposedly brass bell. However, the resulting frequencies from the experimental data and the analytical resolution were very different (more than 1000 Hz difference). Knowing that the model was correct showed that the cycling bell was not made of brass. After verifying the texture and the weight, the bell did not seem to have brass. This model could be used to check the material of a given object.

\subsection{Finite Difference Method}
We have analyzed the theory and how it reflects reality but we have not yet tested the FDM to simulate numerically the frequency heard. We study the bell without damping. Thus, a bell obeys the plate equation (equation \ref{eq:plate}).
\begin{equation}
    \frac{\partial^2 u}{\partial t^2} = -\frac{D}{\rho h}\nabla^4 u
    \label{eq:plate}
\end{equation}
where:
\begin{itemize}
    \item \(\nabla^4\): biharmonic operator (double Laplacian).
\end{itemize}
Boundary conditions are defined as follows:
\begin{itemize}
    \item At the point \(\theta = 0\), vibration is always zero:
    \begin{equation}
         \forall t \ \forall \varphi \ u(0, \varphi, t) = 0 
         \label{eq:conditionfix}
    \end{equation}
    \item Along the great circle \(\theta = \frac{\pi}{2}\), vibration is free along \(\theta\):
    \begin{equation}
        \forall t \ \forall \varphi \ \frac{\partial u}{\partial \theta}(0, \varphi, t) = 0 
        \label{eq:conditionlibre}
    \end{equation}
\end{itemize}
In spherical coordinates, the Laplacian is expressed as:
\begin{equation}
\Delta u = \frac{1}{r^2}\frac{\partial}{\partial r}\left( r^2 \frac{\partial u}{\partial r} \right) + \frac{1}{r^2 \sin \theta}\frac{\partial}{\partial \theta}\left( \sin \theta \frac{\partial u}{\partial \theta} \right) + \frac{1}{r^2 \sin^2 \theta}\frac{\partial^2 u}{\partial \varphi^2}
\label{eq:laplaciensimple}
\end{equation}
Since the shell is very thin, we can consider \(r\) as constant and equal to \(R\). Thus, the waves described by equation \ref{eq:plate} primarily propagate along the surface of the bell, with variations in \(\theta\) and \(\varphi\): \(u = u(\theta, \varphi, t)\).
The Laplacian (equation \ref{eq:laplaciensimple}) becomes:
\begin{equation}
    \frac{1}{R^{2}}\left( \cot\theta\ \frac{\partial u}{\partial\theta} + \frac{\partial^{2}u}{\partial\theta^{2}} + \frac{1}{\left( \sin\theta \right)^{2}}\frac{\partial^{2}u}{{\partial\varphi}^{2}} \right)
\end{equation}
For the double Laplacian, we can factor by\( \frac{1}{R^{4}} \) in the same way and have 9 terms:
\begin{align}
&
    \cot\theta\ \frac{\partial}{\partial\theta}\left( \cot\theta\ \frac{\partial u}{\partial\theta} \right) = \frac{- \cot\theta}{\left( \sin\theta \right)^{2}}\frac{\partial u}{\partial\theta} + \left( \cot\theta \right)^{2}\frac{\partial^{2}u}{\partial\theta^{2}}
    \label{eq:part0}
    \\&
    \cot\theta\ \frac{\partial}{\partial\theta}\left( \frac{\partial^{2}u}{\partial\theta^{2}} \right) = \cot\theta\ \frac{\partial^{3}u}{{\partial\theta}^{3}}
    \label{eq:part1}
    \\&
    \cot\theta\ \frac{\partial}{\partial\theta}\left( \frac{1}{\left( \sin\theta \right)^{2}}\frac{\partial^{2}u}{{\partial\varphi}^{2}} \right) = - 2\frac{\left( \cot\theta \right)^{2}}{\left( \sin\theta \right)^{2}}\frac{\partial^{2}u}{{\partial\varphi}^{2}} + \ \frac{\cot\theta}{\left( \sin\theta \right)^{2}}\frac{\partial^{3}u}{{\partial\theta\partial\varphi}^{2}}
    \label{eq:part2}
    \\&
    \frac{\partial^{2}}{\partial\theta^{2}}\left( \cot\theta\ \frac{\partial u}{\partial\theta} \right) = \frac{2\cot\theta}{\left( \sin\theta \right)^{2}}\frac{\partial u}{\partial\theta} + 2\frac{- 1}{\left( \sin\theta \right)^{2}}\frac{\partial^{2}u}{\partial\theta^{2}} + \cot\theta\frac{\partial^{3}u}{{\partial\theta}^{3}}
    \label{eq:part3}
    \\&
    \frac{\partial^{2}}{\partial\theta^{2}}\left( \frac{\partial^{2}u}{\partial\theta^{2}} \right) = \frac{\partial^{4}u}{\partial\theta^{4}}
    \label{eq:part4}
    \\&
    \frac{\partial^{2}}{\partial\theta^{2}}\left( \frac{1}{\left( \sin\theta \right)^{2}}\frac{\partial^{2}u}{{\partial\varphi}^{2}} \right) = \left( \frac{2}{\left( \sin\theta \right)^{4}} + 4\frac{\left( \cot\theta \right)^{2}}{\left( \sin\theta \right)^{2}} \right)\frac{\partial^{2}u}{{\partial\varphi}^{2}} \notag
    \\ &
    + \frac{- 4\cot\theta}{\left( \sin\theta \right)^{2}}\frac{\partial^{3}u}{{\partial\theta\partial\varphi}^{2}} + \frac{1}{\left( \sin\theta \right)^{2}}\frac{\partial^{4}u}{{\partial\varphi}^{2}{\partial\theta}^{2}}
    \label{eq:part5}
    \\&
    \frac{1}{\left( \sin\theta \right)^{2}}\frac{\partial^{2}}{{\partial\varphi}^{2}}\left( \cot\theta\ \frac{\partial u}{\partial\theta} \right) = \frac{\cot\theta}{\left( \sin\theta \right)^{2}}\frac{\partial^{3}u}{{\partial\theta\partial\varphi}^{2}}
    \label{eq:part6}
    \\
   & \frac{1}{\left( \sin\theta \right)^{2}}\frac{\partial^{2}}{{\partial\varphi}^{2}}\left( \frac{\partial^{2}u}{\partial\theta^{2}} \right) = \frac{1}{\left( \sin\theta \right)^{2}}\frac{\partial^{4}u}{{\partial\varphi}^{2}{\partial\theta}^{2}}
   \label{eq:part7}
    \\
    &\frac{1}{\left( \sin\theta \right)^{2}}\frac{\partial^{2}}{{\partial\varphi}^{2}}\left( \frac{1}{\left( \sin\theta \right)^{2}}\frac{\partial^{2}u}{{\partial\varphi}^{2}} \right) = \frac{1}{\left( \sin\theta \right)^{4}}\frac{\partial^{4}u}{{\partial\varphi}^{4}}
    \label{eq:part8}
\end{align}
We thus obtain with equation \ref{eq:part0} to \ref{eq:part8} the following term:

\begin{align}
    \nabla^{4}u &= \frac{\partial^{4}u}{\partial\theta^{4}} + \frac{1}{\left( \sin\theta \right)^{4}}\frac{\partial^{4}u}{{\partial\varphi}^{4}} + \left( \frac{2}{\left( \sin\theta \right)^{2}} \right)\frac{\partial^{4}u}{{\partial\varphi}^{2}{\partial\theta}^{2}} - 2\frac{\cot\theta}{\left( \sin\theta \right)^{2}}\frac{\partial^{3}u}{{\partial\theta\partial\varphi}^{2}}  \notag 
    \\
    &
    + 2\cot\theta\ \frac{\partial^{3}u}{{\partial\theta}^{3}}
    + \left( \left( \cot\theta \right)^{2} + \frac{- 2}{\left( \sin\theta \right)^{2}} \right)\frac{\partial^{2}u}{\partial\theta^{2}} \notag 
    \\
    & + \left( \frac{2}{\left( \sin\theta \right)^{4}}+ 2\frac{\left( \cot\theta \right)^{2}}{\left( \sin\theta \right)^{2}} \right)\frac{\partial^{2}u}{{\partial\varphi}^{2}} + \frac{\cot\theta}{\left( \sin\theta \right)^{2}}\frac{\partial u}{\partial\theta}
\end{align}
This means:
\begin{align}
    \nabla^{4}u &= \frac{\partial^{4}u}{\partial\theta^{4}} + \frac{\partial^{4}u}{\left( \sin\theta\partial\varphi \right)^{4}} + 2\frac{\partial^{4}u}{\left( \sin\theta\partial\varphi \right)^{2}{\partial\theta}^{2}} \notag 
    \\
    & - 2\cot\theta\frac{\partial^{3}u}{{\partial\theta\left( \sin\theta\partial\varphi \right)}^{2}} + 2\cot\theta\ \frac{\partial^{3}u}{{\partial\theta}^{3}} + \left( \frac{\left( \cos\theta \right)^{2} - 2}{\left( \sin\theta \right)^{2}} \right)\frac{\partial^{2}u}{\partial\theta^{2}} 
    \notag 
    \\
    & 
    + 2\left( \frac{{1 + \left( \cos\theta \right)}^{2}}{\left( \sin\theta \right)^{2}} \right)\frac{\partial^{2}u}{\left( \sin\theta\partial\varphi \right)^{2}} + \frac{\cot\theta}{\left( \sin\theta \right)^{2}}\frac{\partial u}{\partial\theta}
\end{align}
Even:
\begin{align}
    \nabla^{4}u &= \frac{\partial^{4}u}{\partial\theta^{4}} + 2\cot\theta\ \frac{\partial^{3}u}{{\partial\theta}^{3}} + \left( \frac{\left( \cos\theta \right)^{2} - 2}{\left( \sin\theta \right)^{2}} \right)\frac{\partial^{2}u}{\partial\theta^{2}} + \frac{\cot\theta}{\left( \sin\theta \right)^{2}}\frac{\partial u}{\partial\theta}
    \notag 
    \\
    &
    + \frac{\partial^{4}u}{\left( \sin\theta\partial\varphi \right)^{4}} + 2\left( \frac{{1 + \left( \cos\theta \right)}^{2}}{\left( \sin\theta \right)^{2}} \right)\frac{\partial^{2}u}{\left( \sin\theta\partial\varphi \right)^{2}}
    \notag 
    \\
    &
    + 2\frac{\partial^{4}u}{\left( \sin\theta\partial\varphi \right)^{2}{\partial\theta}^{2}} - 2\cot\theta\frac{\partial^{3}u}{{\partial\theta\left( \sin\theta\partial\varphi \right)}^{2}}
    \label{eq:secondlapsimplifyed}
\end{align}
Remember that for ~\(u(\theta,\varphi)\):
\begin{align*}
    \varphi &\in \left\lbrack 0;2\pi \right\lbrack\ \ \ \ ( + 2k\pi)
    \\
    \theta &\in \left\lbrack 0;\frac{\pi}{2} \right\rbrack
\end{align*}
We can cut in small pieces and get \(u(\theta,\varphi) = u(n,k)\), this approximates the terms of equation \ref{eq:secondlapsimplifyed} :
\begin{align}
    \frac{\partial u}{\partial\theta}(n,k) &\cong \frac{1}{2\delta\theta}\left( u(n + 1,k) - u(n - 1,k) \right) \label{eq:simpl0} \\
    \frac{\partial^{2}u}{{\partial\theta}^{2}}(n,k) &\cong \frac{1}{{\delta\theta}^{2}}\left( u(n + 1,k) - 2u(n,k) + u(n - 1,k) \right) \label{eq:simpl1} \\
    \frac{\partial^{2}u}{{\partial\varphi}^{2}}(n,k) &\cong \frac{1}{{\delta\varphi}^{2}}\left( u(n,k + 1) - 2u(n,k) + u(n,k - 1) \right) \label{eq:simpl2} \\
    \frac{\partial^{3}u}{{\partial\theta}^{3}}(n,k) &\cong \frac{3}{{8\delta\theta}^{3}} \Big( u(n + 2,k) - u(n - 2,k) \nonumber \\
    &\quad - 2\left( u(n + 1,k) - u(n - 1,k) \right) \Big) \label{eq:simpl3} \\
    \frac{\partial^{4}u}{{\partial\theta}^{4}}(n,k) &\cong \frac{1}{{\delta\theta}^{4}} \Big( u(n + 2,k) + u(n - 2,k) - 4u(n + 1,k) \nonumber \\
    &\quad - 4u(n - 1,k) + 6u(n,k) \Big) \label{eq:simpl4}\\
    \frac{\partial^{4}u}{{\partial\varphi}^{4}}(n,k) &\cong \frac{1}{{\delta\varphi}^{4}} \Big( u(n,k + 2) + u(n,k - 2) - 4u(n,k + 1) \nonumber \\
    &\quad - 4u(n,k - 1) + 6u(n,k) \Big) \label{eq:simpl5} \\
    \left( \frac{\partial^{3}u}{{\partial\varphi\partial\theta}^{2}} \right)(n,k) &\cong \frac{1}{2\delta\varphi\, {\delta\theta}^{2}} \Big( u(n + 1,k + 1) - 2u(n,k + 1) + u(n - 1,k + 1) \nonumber \\
    &\quad - u(n + 1,k - 1) + 2u(n,k - 1) - u(n - 1,k - 1) \Big) \label{eq:simpl6} \end{align}
    \begin{align}
    \left( \frac{\partial^{3}u}{{\partial\theta\partial\varphi}^{2}} \right)(n,k) &\cong \frac{1}{2{\delta\varphi}^{2}\delta\theta} \Big( u(n + 1,k + 1) - 2u(n + 1,k) + u(n + 1,k - 1) \nonumber \\
    &\quad - u(n - 1,k + 1) + 2u(n - 1,k) - u(n - 1,k - 1) \Big) \label{eq:simpl7} \\
    \left( \frac{\partial^4 u}{\partial \theta^2 \partial \varphi^2} \right)(n, k) &\cong  
    \frac{1}{\delta \varphi^2 \delta \theta^2} \Big(
    u(n+1, k+1) - 2u(n, k+1) + u(n-1, k+1) \nonumber \\
    &\quad - 2\left( u(n+1, k) - 2u(n, k) + u(n-1, k) \right) \nonumber \\
    &\quad + u(n+1, k-1) - 2u(n, k-1) + u(n-1, k-1) \Big) \label{eq:simpl8}
\end{align}
However, to consider the time, we need to add the damping term:
\begin{equation}
    \rho h\frac{\partial^{2}u}{{\partial t}^{2}} = - \frac{D}{R^{4}}\ \nabla^{4}u - \sigma\frac{\partial u}{\partial t}
\end{equation}
We also know that:
\begin{equation}
\frac{\partial^{2}u}{{\partial t}^{2}}(n,k,t) \cong \frac{1\ }{{\delta t}^{2}}\left( u(n,k,t + 1) - 2\ u(n,k,t) + u(n,k,t - 1) \right)
    \label{eq:timeequiv2}
\end{equation}
So using equation \ref{eq:timeequiv2} in \ref{eq:plate} we get:

\begin{align}
    &\left( \frac{\rho h\ }{{\delta t}^{2}} + \frac{\sigma}{2\delta t} \right)u(n,k,t + 1) = - \frac{D}{R^{4}}\ \nabla^{4}u(n,k,t) 
    \nonumber
    \\ &+ 2\ \frac{\rho h\ }{{\delta t}^{2}}u(n,k,t) 
    + \left( - \frac{\rho h\ }{{\delta t}^{2}} + \frac{\sigma}{2\delta t} \right)u(n,k,t - 1)
\end{align}
For the condition at the limits equation \ref{eq:conditionfix}, and \ref{eq:conditionlibre} we get: 
\begin{itemize}
\item
  \(\begin{matrix}
  \forall t\ \ \ \forall\varphi\ \ \ \ u(0,\varphi,t) = 0 \\
  \end{matrix}\)
\item
  \(\begin{matrix}
  \forall t\ \ \ \forall\varphi\ \ \ \ \frac{\partial u}{\partial\theta}(\frac{\pi}{2},\varphi,t) = 0 \\
  \end{matrix}\)
\end{itemize}

The first implemented code in Python was the FDM for the bell for the equation \ref{eq:notgood} that does not give audible frequencies. We assumed that the mechanical vibrations in the material obey D’Alembert’s wave
equation. This is not an unreasonable assumption since it is known to hold for
waves in fluids, i.e., longitudinal waves. 
\begin{equation}
\Delta u - \frac{1}{c^2}\frac{\partial^2 u}{\partial t^2} = 0 
\label{eq:notgood}
\end{equation}
 The time of the simulation could go up to four hours. The simulation time of the code for the equation \ref{eq:plate} developed previously would be even greater. In Matlab, the calculations are time-consuming for equation \ref{eq:plate} and require very small steps to converge. The steps were so small that, because of the lack of time, no conclusion could be made. To conclude, in Python, further implementation is required. A draft was started, but computation time is still concerning.
All in all, it was demonstrated that equation \ref{eq:plate} is a reliable model. 

\section{Conclusion}
\subsection{General conclusion}

When comparing FEM and FDM relative errors in frequency measurement when changing the tension, the time interval or the number of nodes, there was a relationship between the periodicity of the error and the degree of the harmonic. This relationship is probably due to the discretizations done in each method. Moreover, in the case of the tension change, between the periodicity of the error in FEM and FDM at a given harmonic or at the fundamental, a phase shift of $\pi /2 $ is visible. In the two other cases of changing the time interval or the number of nodes, both FEM and FDM have the same phase. 

The error associated with frequency measurements is negligible in most practical applications.  The models discussed are primarily intended for comparison with real-world conditions, where measurement instruments impose inherent limitations. For example, the fundamental frequency of a waveform is typically captured using a standard microphone with a sampling rate of 44100 Hz. This sampling rate directly influences the resolution of the FFT, thereby constraining the accuracy of the extracted frequency. Similarly, when simulating a waveform, the choice of sampling rate introduces a corresponding measurement uncertainty. As demonstrated in this study, the discrepancy between frequencies obtained through the FFT, whether from simulated or experimental data and those derived analytically, is limited to a few hertz. Notably, despite this deviation from the theoretical solution, the measurement error observed in simulations remains representative of real-world behavior, as similar margins of error are consistently observed in experimental contexts.

Computation time depends significantly on the implementation. However, for this study, the computational time for both methods is approximately equivalent. While the FDM is generally faster, the FEM tends to be more efficient when handling a higher number of nodes. Additionally, FDM converges for larger time intervals compared to FEM, indicating that FDM reaches a stable solution "more quickly" than FEM.

From a mathematical and implementation perspective, FDM is more straightforward but requires knowledge of the object’s shape, necessitating recalculations for different geometries. In contrast, while FEM is conceptually more complex to understand and implement, once the initial derivations are established, the object type and boundary conditions are incorporated into the stiffness (K) and mass (M) matrices. This allows for a more generalized approach, where a single FEM algorithm can be adapted for different objects simply by modifying these matrices.

Furthermore, given the availability of open-source FEM software, it is evident that in an industrial context, FEM would be more cost-effective and flexible. Developing a single general FEM algorithm and adapting it by modifying the K and M matrices would be more efficient for large-scale applications requiring frequent modifications. Conversely, FDM remains the more practical choice for smaller and more specialized applications due to its relative simplicity and ease of implementation. However, as experience with the bell, coding FDM can quickly become a hassle because of the long equations.

\subsection{Additional possible work}
Further implementation and investigation are recommended. A formal comparison of the algorithmic complexity of FEM and FDM would provide clearer insights into computational efficiency. Implementing FEM for the bell model using Python or an open-source FEM simulator would yield practical information regarding implementation challenges \cite{konate1986}. Additionally, examining the frequencies of guitar strings by incorporating the modulus of elasticity could demonstrate its negligible impact on frequency results. Finally, exploring alternative numerical methods such as the Boundary Element Method (BEM) would highlight the existence of methods beyond FEM and FDM.

\end{document}